\documentclass[journal]{IEEEtran}

\usepackage{amsmath}

\usepackage{cite}
\usepackage{algorithm}
\usepackage{algpseudocode}
\usepackage{enumerate}
\usepackage{color}
\usepackage{setspace}
\usepackage{balance}
\usepackage{url}
\usepackage{stfloats}

\usepackage[pdftex]{graphicx}
\graphicspath{{./}}%
\DeclareGraphicsExtensions{.jpg,.png,.pdf}
\usepackage[font={footnotesize}]{caption}
\usepackage[caption=false,font=footnotesize]{subfig}

\newcommand{\bi}{\begin{itemize}}	
\newcommand{\ei}{\end{itemize}}
\newcommand{\bn}{\begin{enumerate}}	
\newcommand{\en}{\end{enumerate}}
\newcommand{\bc}{\begin{center}}
\newcommand{\ec}{\end{center}}
\newcommand{\be}{\begin{equation}}
\newcommand{\ee}{\end{equation}}
\newcommand{\bea}{\begin{eqnarray}}
\newcommand{\eea}{\end{eqnarray}}
\newcommand{\ben}{\begin{equation*}}
\newcommand{\een}{\end{equation*}}
\newcommand{\beqa}{\begin{eqnarray}}
\newcommand{\eeqa}{\end{eqnarray}}

\newcommand{\mb}{\mathbf}

\begin{document}
\title{Target Tracking in Confined Environments with Uncertain Sensor Positions}

\author{Vladimir~Savic, Henk~Wymeersch, and Erik~G.~Larsson
\thanks{Copyright (c) 2015 IEEE. Personal use of this material is permitted. However, permission to use this material for any other purposes must be obtained from the IEEE by sending a request to pubs-permissions@ieee.org. The original version of this manuscript is published in IEEE Transactions on Vehicular Technology (DOI: 10.1109/TVT.2015.2404132).}%
\thanks{ V. Savic and E. G. Larsson are with the Dept. of Electrical Engineering (ISY), Link\"{o}ping University, Sweden (e-mails: vladimir.savic@liu.se, erik.g.larsson@liu.se). H. Wymeersch is with the Dept. of Signals and Systems, Chalmers University of Technology, Gothenburg, Sweden (e-mail: henkw@chalmers.se).}%
\thanks{The work was supported by the project Cooperative Localization (CoopLoc), funded by the Swedish Foundation for Strategic Research (SSF); ELLIIT; Swedish Research Council (VR), under grant no. 2010-5889; and European Research Council, under grant COOPNET No. 258418. Parts of this work were presented at the Intl. Conf. on Information Fusion, 2013 \cite{Savic2013fusion}.}%
}

\maketitle

\begin{abstract}
To ensure safety in confined environments such as mines or subway tunnels, a (wireless) sensor network can be deployed to monitor various environmental conditions. One of its most important applications is to track personnel, mobile equipment and vehicles. However, the state-of-the-art algorithms assume that the positions of the sensors are perfectly known, which is not necessarily true due to  imprecise placement and/or dropping of sensors. Therefore, we propose an automatic approach for simultaneous refinement of sensors' positions and target tracking. We divide the considered area in a finite number of cells, define dynamic and measurement models, and apply a discrete variant of belief propagation which can efficiently solve this high-dimensional problem, and handle all non-Gaussian uncertainties expected in this kind of environments. Finally, we use ray-tracing simulation to generate an artificial mine-like environment and generate synthetic measurement data. According to our extensive simulation study, the proposed approach performs significantly better than standard Bayesian target tracking and localization algorithms, and provides robustness against outliers.
\end{abstract}

\section{Introduction}\label{sec:intro}

\subsection{Background and Motivation}

A confined environment represents a constrained and irregularly-shaped area, consisting of a series of tunnels or passages that connect different rooms or halls. Typical examples are underground mines, caves, steel factories and subways. In these environments, the working conditions may be hazardous due to the possibilities of traffic accidents, machine collisions, wall collapses, fires and explosions. These environments require continuous monitoring using sensors deployed all over the area. The sensors may be wired and connected to  control rooms, but to improve the safety and reduce operational costs, recently the industry is developing  robust wireless communication systems for   this kind of environments \cite{Stenumgaard2013,Misra2010,Yarkan2009}.

A wireless sensor network (WSN) can be deployed across the area to monitor the environmental conditions such as stability, temperature and gas levels. The information obtained from the sensors can be used to control the ventilation system, and determine the unsafe areas and rescue paths. Beyond this ability, a WSN can be used to track the personnel, mobile equipment and vehicles. The problem is very challenging due to the unavailability of GPS signals and the characteristics of the  propagation environment. The knowledge of the last location of an employee is especially important in the aftermath of accidents such as a wall collapse, explosion, or water inundation, but can be also used for task optimization, production monitoring and traffic management. For instance, according to the MINER act \cite{MINER2006}, created in response to the many mine tragedies in the United States during 2006, the emergency response plan ``shall provide for above-ground personnel to determine the current or immediately pre-accident location of all underground personnel''. This problem, that also exists in many other confined environments, is the main motivation behind the work reported in this paper.

\subsection{Related Work}\label{subsec:rel} 

Contemporary techniques for localization and tracking in confined environments are very basic. They are typically based on manual reporting of the employee's location using paging phones or video surveillance \cite{Chehri2009,Novak2009}. Moreover, there are few proposals in the literature, based on fingerprinting \cite{Nerguizian2006, Dayekh2010,Lim2013,Lee2011}, trilateration \cite{Chehri2009,Chehri2012}, centroid \cite{Li2009} and Bayesian filtering \cite{Hedley2013, Savic2014tw}.

More specifically, in \cite{Nerguizian2006}, a fingerprinting technique was proposed, in which seven relevant parameters (including mean excess delay, total received power, and delay spread) were learned offline from   wideband  impulse responses measured at hundreds of  locations. Then, these 7D vectors were used as the input to an artificial neural network pattern-matching algorithm. The measurements were conducted in a gallery of the CANMET mine, a former gold mine located in Quebec, Canada. This method was then improved in \cite{Dayekh2010} by using more receivers with known positions. The fingerprinting techniques \cite{Lim2013, Lee2011}, based on WiFi signals, have been also applied in subway tunnels in Seoul, S.\ Korea. The main problem of these algorithms is that they are not well suited for dynamic propagation environments (e.g., caused by movement of heavy machinery) in which the fingerprints have to be updated very frequently.

In \cite{Chehri2009,Chehri2012}, ultra-wideband (UWB) measurements were used for positioning. They were motivated by a high ranging accuracy in cluttered environments and low-cost implementation of the communication system. To solve the trilateration problem, many types of algorithms have been applied, including linearized least-squares, Gauss-Newton and bounding-box methods. The measurements were performed in the same environment as the one studied in \cite{Nerguizian2006}. The main drawbacks of these algorithms are that the sensor nodes have to be precisely deployed and maintained, and that the algorithms are sensitive to outliers.

In \cite{Li2009}, a centroid algorithm was proposed, in which the miner's location was found by averaging the coordinates of the detected anchors. The algorithm is a part of a structure-aware self-adapting (SASA) WSN, which is capable of detecting structure variations caused by mine collapses. The main problem of this approach is that it requires a high density of uniformly deployed sensor nodes.

In \cite{Hedley2013}, Bayesian point-mass (grid-based) filtering was applied to track mine vehicles. The main goal was monitoring and control of ore extraction from the draw points in a mine in Australia. Since the draw points are very close to each other, high tracking accuracy is required. The main problem of this approach is that it requires many grid points in order to obtain sufficiently accurate estimates.

Finally, the results of the measurement campaign \cite{Savic2014tw}, carried out in a basement tunnel of Link\"oping university and an iron-ore mine in Kiruna, Sweden, indicated that UWB time-of-arrival (TOA) allows very accurate ranging in line-of-sight (LOS) and non-LOS (NLOS) scenarios caused by thin obstacles. However, if the direct path is blocked by a thick tunnel wall, the TOA-based ranging leads to a relatively large bias. Moreover, the analysis showed that NLOS conditions cannot be accurately discriminated from LOS conditions,  which means that (Bayesian) soft-decision algorithms  are required for accurate ranging and positioning in these environments.
 
The previously described state-of-the-art algorithms assume that the positions of the sensors are perfectly known, which is not necessarily the case due to imprecise placement and/or sensor drops caused by vibrations or wall collapses.\footnote{Although probably not available in confined environments nowadays, we also envision that uncertain sensors' positions can be an outcome of some (cooperative) sensor network localization algorithm \cite{Patwari2005,Wymeersch2012}.} One possible solution to this problem is to  manually and periodically verify that the sensors positions are correct. However, this approach may be too costly and even infeasible in some areas due to the on-going activities.

\subsection{Technical Contributions}\label{subsec:contr}

In this paper we propose an automatic approach for target tracking with uncertain sensor positions, which involves both simultaneous refinement of the sensors position estimates (localization) and target tracking (SLAT). Our specific technical contributions are as follows:
\bi
\item We divide the considered area into a finite number of cells, and define appropriate dynamic and TOA measurement models that take into account the quantization effects
associated with this division.
\item We formulate the localization and target tracking problem in a Bayesian setting and apply a discrete variant of belief propagation (BP). The resulting proposed algorithm (referred to as SLAT-BP) can  efficiently  handle the high dimensionality of the problem and the  non-Gaussian uncertainties.
\item To demonstrate the performance of our SLAT-BP algorithm, we perform an extensive simulation study using synthetic impulse responses obtained from ray-tracing simulation of a   mine-like environment.  Our results show that SLAT-BP performs significantly better than standard Bayesian target tracking and localization algorithms, and provides robustness against outliers.
\ei

\subsection{Paper organization}\label{subsec:org} 

The remainder of this paper is organized as follows. In Section~\ref{sec:sys-mod}, we formulate the problem and define the dynamic and measurement models. In Section \ref{sec:slat}, we propose the algorithm for simultaneous localization and tracking, based on real-time belief propagation. TOA error modelling using ray-tracing simulation and performance analysis are provided in Section~\ref{sec:sims}. Finally, conclusions and proposals for future work are provided in Section~\ref{sec:conc}.

\section{System model}\label{sec:sys-mod} 

\subsection{Problem formulation}\label{subsec:problem} 

\begin{figure}[!t]
\includegraphics[width=1.06\columnwidth]{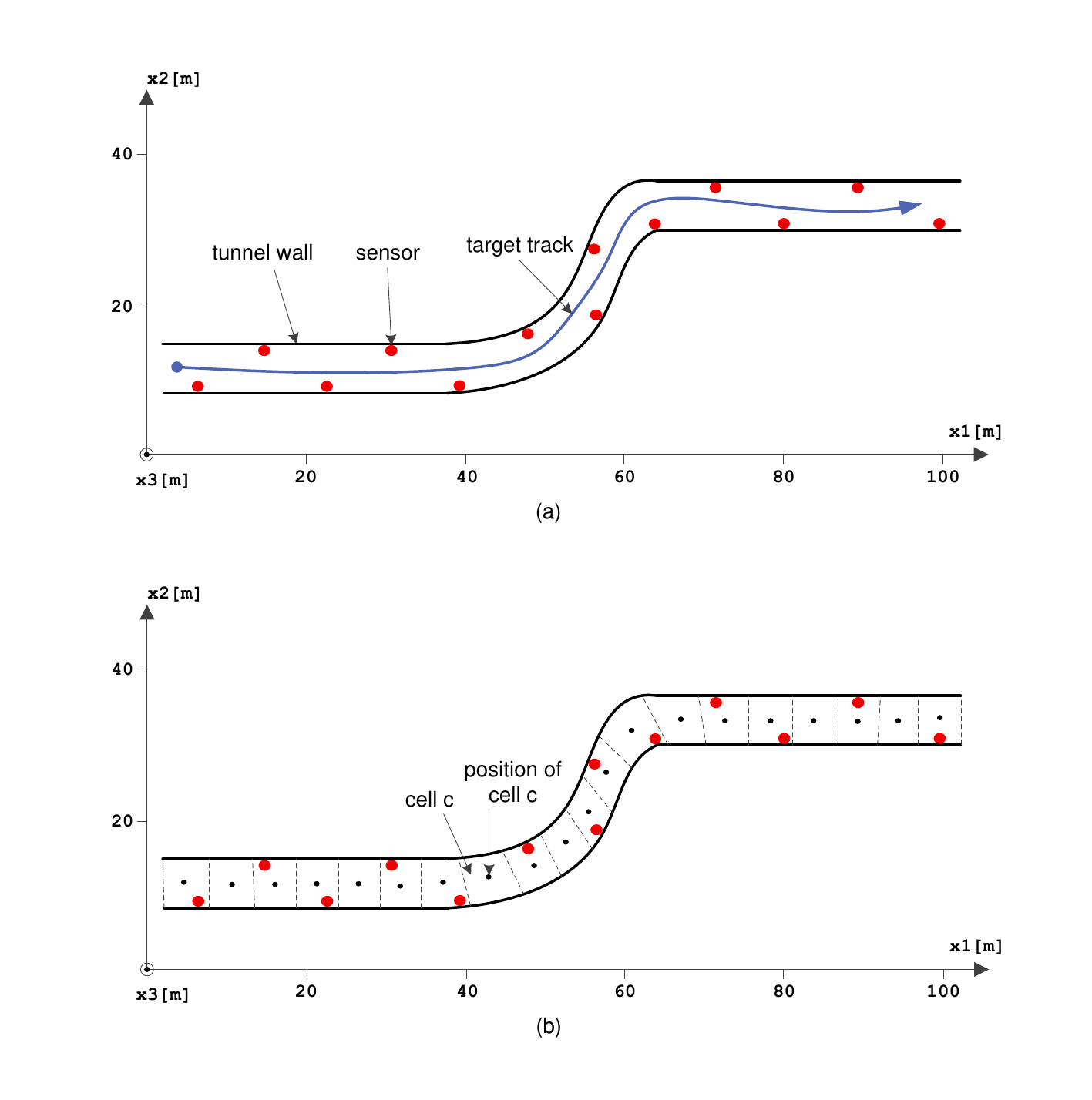}
\caption{Illustration of target tracking in a confined environment: (a) deployment of the sensors and a target track, (b) division of the area into 20 cells.}
\label{fig:mine-cells}
\end{figure}

We consider $N_s$ sensors with fixed 3D positions $\mb{z}_n=(z_{n,1},z_{n,2},z_{n,3})$, $n=1,2,\ldots,N_s$, and one target, with 3D position $\mb{x}_t=(x_{t,1},x_{t,2},x_{t,3})$, at time $t$, $t=1,2,\ldots,N_T$, moving through the confined area. Fig. \ref{fig:mine-cells}a illustrates the scenario. The sensors are usually placed on the walls or the ceiling, but their positions are not perfectly known. A moving target periodically emits a signal (including a unique identifier)\footnote{That means that our algorithm can be also used for multi-target tracking, simply by running the same algorithm multiple times. Otherwise, different algorithms, e.g., with data association \cite{Cetin2006}, would be necessary.} that can be detected by a subset of the sensors, with a sampling interval $T_s$. The target is equipped with an inertial measurement unit (IMU), so it periodically communicates its measured velocity. We also assume that there are one or more fusion centers (FCs) (e.g., a computer in a control room or a target itself), which have available the priors of the sensors' and target's positions, and periodically collect  measurements from the sensors and the target. 

A confined environment is naturally a  continuous 3D space, but typically it is irregular and it is impossible to analytically describe the shape of its borders. On the other hand, using an unrestricted 3D continuous space would lead to decreased computational efficiency, and more importantly, significant loss of performance in that  position estimates could end up, for example,   behind a wall. Therefore, we propose to use a   \textit{discrete} 3D space, in which the environment is divided into a finite number of cells. The 3D position of the cell $\mb{l}_c=(l_{c,1},l_{c,2},l_{c,3})$ ($c=1,...,N_c$, where $N_c$ is the number of cells) is represented, in Cartesian coordinates, by the approximation of its geometrical center. It is thereby assumed that the FCs have available a detailed floor plan of the whole area. The cell size must be chosen based on a  trade-off between computational complexity and performance, and it is preferable that all the cells have approximately the same size. With this model, in which $\mb{z}_n$ and $\mb{x}_t$ are discrete variables, our goal is to identify in which cells the target and sensors are located. This approach  also facilitates the application of belief propagation (see Section \ref{subsec:slat-bp}) without applying Monte Carlo or other approximations. Fig. \ref{fig:mine-cells}b illustrates a confined environment divided into cells.

Finally, we assume that the prior knowledge of the sensors and the target positions is defined by probability mass functions (PMFs) $p_{n,0}({\bf{z}}_{n,0})$, and $p_{0}({\bf{x}}_{0})$. While the target's prior represents information about its initial position, the priors of the sensors' positions represent the available information on the real positions of the sensors. In practice, this information may be   provided by the deployment team, and may   be imperfect due to  imprecise deployment or drops of sensors.

\subsection{Dynamic Model based on IMUs}\label{subsec:dyn} 

We assume that the target is equipped by an IMU,  consisting of a 3-axis accelerometer, and a 3-axis gyroscope, that aids the tracking. IMUs are relatively insensitive to environmental conditions, and a good choice in the motivating applications. The IMU periodically provides (each  $T_s$ second) a velocity estimate in all 3 directions $\mb{v}_{t}=(v_{t,1},v_{t,2},v_{t,3})$. The coordinate in the $\kappa$-th dimension ($\kappa=1,2,3$) of the measured velocity at time $t$ can be written as a function of the target's positions as:
\beqa\label{eq:velocity-slat} 
v_{t,\kappa}=\frac{x_{t,\kappa}-x_{t-1,\kappa}}{T_s}+u_{t,\kappa}^q+u_{t,\kappa}^m,
\eeqa 
\ben
~~~u_{t,\kappa}^q \sim {p_q}({u^q})={\rm{Unif}}(u^q;-\frac{D}{T_s},\frac{D}{T_s}),
\een
\ben
u_{t,\kappa}^m \sim {p_m}({u^m}) ={\mathcal{N}}(u^m;0,{\sigma_u^2}),
\een
where $u_{t,\kappa}^q$ and $u_{t,\kappa}^m$ are samples (at time $t$, and dimension $\kappa$) of the quantization and measurement noise, drawn from the uniform ${\rm{Unif}}(\cdot)$, and the Gaussian distribution ${\mathcal{N}}(\cdot)$, respectively. The parameter $\sigma_u^2$ represents the variance of the measurement noise of the IMU, and $-D/T_s$, $D/T_s$ represent the boundaries of the uniform distribution. The quantization noise is introduced into the model here because we have to relate the continuous quantity $v_{t,\kappa}$ with the discrete quantity $(x_{t,\kappa}-x_{t-1,\kappa})$. Since the boundaries of the irregularly-shaped cell are difficult to compute, we approximated the cell with a circumscribed cube with length $D$ ($D=\max\limits_{c,\kappa}{D_{\kappa,c}}$, where $D_{\kappa,c}$ is the maximum possible distance over dimension $\kappa$ in cell $c$). Note that as the cell size $D$ decreases, the   quantization noise will eventually vanish.

The distribution of the total IMU noise $u=u^q+u^m$ is given by following convolution: 
\beqa\label{eq:conv-process} 
&{p_u}(u) &= \int {{p_m}(u - {u^q}){p_q}({u^q})d{u^q}}\\
&~& \propto {\Phi}(u + \frac{D}{T_s}) - {\Phi}(u - \frac{D}{T_s})\nonumber,
\eeqa
where $\Phi(u)=0.5 \left(1+\mathrm{erf}(u/(\sigma_u\sqrt{2})\right)$ is the cumulative Gaussian distribution. Now, we can model the dynamics\footnote{In some of the literature, the quantity $p(\mb{v}_{t},{{\bf{x}}_t}|{{\bf{x}}_{t - 1}})$ is denoted by  $p({\bf{x}}_t|{\bf{x}}_{t-1})$. We prefer to  write out the measured velocity explicitly.} of the target as:
\beqa\label{eq:dynamic-slat}
&p(\mb{v}_{t},{{\bf{x}}_t}|{{\bf{x}}_{t - 1}}) &= p(\mb{v}_t|{\bf{x}}_t,{\bf{x}}_{t-1})p({\bf{x}}_t|{\bf{x}}_{t-1}) \\
&~& \propto p(\mb{v}_t|{\bf{x}}_t,{\bf{x}}_{t-1})\nonumber \\
&~& = \prod\limits_{\kappa = 1,2,3} {p(v_{t,\kappa}|{x_{t,\kappa}},{x_{t - 1,\kappa}})}\nonumber\\
&~&=\prod\limits_{\kappa = 1,2,3} p_u(v_{t,\kappa}-(x_{t,\kappa}-x_{t-1,\kappa})/T_s)\nonumber,
\eeqa
where we assumed independence between the coordinates, and that the target's mobility model is unknown (i.e., $p({\bf{x}}_t|{\bf{x}}_{t-1}) \propto 1$).\footnote{Even without this assumption, the IMU typically provides much more precise information than the mobility model, so $p(\mb{v}_t|{\bf{x}}_t,{\bf{x}}_{t-1})p({\bf{x}}_t|{\bf{x}}_{t-1}) \approx p(\mb{v}_t|{\bf{x}}_t,{\bf{x}}_{t-1})$.} Recall also that ${{\bf{x}}_t}$ is discrete, so \eqref{eq:dynamic-slat} gives us information about the cell of the target at time $t$, given the cell of the target at time $t-1$, and the measured velocity $\mb{v}_{t}$.

For our framework (see Section \ref{subsec:graph-model}), it is convenient to formally define the ``dynamics'' of the sensors:
\be\label{eq:dynamic-slat2} 
p({{\bf{z}}_{n,t}}|{{\bf{z}}_{n,t - 1}}) = \delta({{\bf{z}}_{n,t}}-{{\bf{z}}_{n,t-1}}),
\ee
where $\delta(\cdot)$ is the Dirac delta impulse, which enforces that the sensors are static (${\bf{z}}_{n,t}={\bf{z}}_{n,t-1}={\bf{z}}_{n}$).

Finally, we note that, if $p_{0}({\bf{x}}_{0})$ is very informative (e.g., the target's initial cell is given), the target's dynamic model (given by \eqref{eq:dynamic-slat}) already constitutes sufficient information for a tracking algorithm known as \textit{dead reckoning} \cite{Jin2013}, but this approach would suffer from error accumulation over time. Therefore, we need to use a WSN which will provide periodic measurements w.r.t.\ their positions.

\subsection{TOA Measurement Model}\label{subsec:measure}

We assume that round-trip TOA (RT-TOA) measurements are obtained, at each time slot, by a subset of the sensors in proximity to the target. We decided to use RT-TOA (instead of one-way TOA), in order to avoid the need for clock synchronization between the sensors and the target. We did not consider received signal strength (RSS) since the distance estimates will be highly erroneous, due to the severe multipath in confined environments, as shown in \cite{Boutin2008,Chehri2012,Chuasomboon2013}. RT-TOA can be obtained using many techniques \cite{Dardari2009}, but we assume that a simple thresholding is performed. More exactly, the RT-TOA is taken to be the arrival time of the first multipath component in the measured impulse response that exceeds a predefined threshold. Note  that a signal should have very large bandwidth (WB or UWB), in order to provide high TOA resolution \cite{Gezici2005}.

The measured distance between sensor $n$ and the target at time $t$, can be written as:
\be\label{eq:dist-toa}
d_{t,n}=c\tau_{t,n}-d_{\mathrm{PT}}=\left\| {\mb{x}_t-\mb{z}_{n,t}} \right\| + w_{t,n}^q+w_{t,n}^m,
\ee
\ben
w_{t,n}^q \sim {p_q}({w^q})={\rm{Unif}}(w^q;0,D\sqrt{3}),~w_{t,n}^m \sim {p_m}({w^m}),
\een
where $d_{\mathrm{PT}}$ is a known bias caused by processing time on a target, $\tau_{t,n}$ is measured TOA, $c=3 \cdot 10^8$ m/s is the speed of light, and $w_{t,n}^q$, $w_{t,n}^m$ are samples of the quantization noise, and measurement noise, respectively. The distribution of the quantization noise is not available in a parametric form, so we choose the least informative (i.e., a uniform) distribution to keep the algorithm tractable.

While the measurement noise depends on many factors (such as thermal noise, bandwidth, and the quality of the sensors), the bias arising from multipath propagation in NLOS conditions is usually the most critical source of error. The most common approach \cite{Marano2010,Miao2007} is to identify NLOS measurements, and discard them, or alternatively mitigate the effect of the multipath bias. However, these techniques require an identification of NLOS conditions, which cannot be always accurately done (see, for example, \cite{Marano2010, Savic2014tw}). Therefore, we prefer a model that does not require NLOS identification, but only knowledge of the probability of having NLOS.

According to previous results \cite{Chehri2009}, we can roughly model line-of-sight (LOS) measurement noise with a Gaussian distribution, and NLOS noise (in which the walls block the direct path) with a Weibull or an exponential distribution. However, in severe multi-path environments, it is expected that NLOS noise has multiple modes (see also Section~\ref{subsec:ray-tracing}). Therefore, we use a Gaussian mixture (GM), which is capable of approximating arbitrary probability distributions. Moreover, we also need to take into account other sources of NLOS, such as vehicles and machinery. Since these objects are usually dynamic, made of different materials, and have different sizes and thicknesses, it is difficult to model  their effect. Therefore, we assume that this error is uniformly distributed \cite{Hedley2013}, but that it appears with very small probability. In total, the model for $p_w(w^m)$ is then given by the following mixture:
\beqa\label{eq:measure-slat}
&{p_m}({w^m}) &= {P_{{\rm{LOS}}}}\cdot{\cal N}({w^m};0,\sigma_{w,0}^2)+\\
&~&{P_{{\rm{NLOS}}}}\cdot\sum\limits_{i=1,\ldots,N_{M}} {{\rho_{w,i}}{\cal N}({w^m};{\mu_{w,i}},\sigma_{w,i}^2)}+\nonumber\\ &~&P_{{\rm{OBS}}}\cdot{\rm{Unif}}(w^m;0,D_{\mathrm{max}})\nonumber,
\eeqa
where $\sigma_{w,0}$ is the standard deviation of the LOS component of the noise; $\rho_{w,i}$, $\mu_{w,i}$, $\sigma_{w,i}$ are the weights, means and standard deviations of the NLOS noise caused by tunnel walls; $N_{M}$ is the number of GM components; and $D_{\mathrm{max}}$ is the maximum distance error. $P_{\mathrm{LOS}}$, $P_{\mathrm{NLOS}}$ and $P_{\mathrm{OBS}}$ are the probabilities of LOS, NLOS caused by tunnel walls, and NLOS caused by other obstacles, respectively (with $P_{\mathrm{LOS}}=1-P_{\mathrm{NLOS}}-P_{\mathrm{OBS}}$).
While $P_{\mathrm{NLOS}}$ and $P_{\mathrm{OBS}}$ can be approximately estimated by examining the floor plan of the deployment area, the GM parameters ($\rho_{w,i}$, $\mu_{w,i}$ and $\sigma_{w,i}$) can be estimated by applying the expectation-maximization (EM), generalized EM, or the k-means algorithm \cite[Chapter 9]{Bishop2006} on training samples.

The distribution of the total noise $w=w^m+w^q$ is given by the following convolution:
\beqa\label{eq:conv-measure}
&{p_w}(w) &= \int {{p_m}(w - {w^q}){p_q}({w^q})d{w^q}}  \\
&~&= {\frac{P_{\mathrm{LOS}}}{D\sqrt{3}}}\left({\Phi_0}(w) - {\Phi_0}(w - D\sqrt{3})\right)+\nonumber\\
&~&{\frac{P_{\mathrm{NLOS}}}{D\sqrt{3}}}\sum\limits_{i=1,\ldots,N_{M}}{{\rho_{w,i}} \left({\Phi_i}(w) - {\Phi_i}(w - D\sqrt{3})\right)}+\nonumber\\
&~&g(w)P_{{\rm{OBS}}}\nonumber,
\eeqa
where ${\Phi_i}(\cdot)$ is the shorthand notation for ${\Phi}(\cdot;{\mu_{w,i}},\sigma_{w,i}^2)$, and the distribution $g(w)$ is found by convolution of two uniform distributions:
\be\label{eq:gw}
g(w) = \left\{ \begin{array}{l}
w/({{D}_{\max }}D\sqrt 3), \,\,\,\,\,\,\,\,\,\,\,\,\,\,\,\,\,\,\,\,\,\,\,\,\,\,\,\,\,\,\,\,0 < w < D\sqrt 3 \\
1/{{D}_{\max }}\,,\,\,\,\,\,\,\,\,\,\,\,\,\,\,\,\,\,\,\,\,\,\,\,\,\,\,\,\,\,\,\,\,\,\,\,\,\,\,\,\,\,\,\,\,\,\,\,\,D\sqrt 3  < w < {{D}_{\max }}\,\\
(D^{'}_{\max }-w)/({{D}_{\max }} D\sqrt3),\,\,\,{D_{\mathrm{max}}<w<D^{'}_{\max }}\\
0\,,\,\,\,\,\,\,\,\,\,\,\,\,\,\,\,\,\,\,\,\,\,\,\,\,\,\,\,\,\,\,\,\,\,\,\,\,\,\,\,\,\,\,\,\,\,\,\,\,\,\,\,\,\,\,\,\,\,\,\,\,\,\,\,\,\,{\rm{otherwise}}
\end{array} \right.
\ee
where $D^{'}_{\max }=D\sqrt3+D_{\max }$. Finally, the likelihood function is given by:
\be\label{eq:lh-slat}
p(d_{n,t}|\mb{x}_t,\mb{z}_{n,t}) = p_w(d_{n,t}-\left\| {\mb{x}_t-\mb{z}_{n,t}} \right\|).
\ee

The likelihood functions can be computed for all sensors which detect the target. However, we assume that the sensors can perform measurements with the target if and only if $d_{n,t}<d_{\mathrm{TH}}$ where $d_{\mathrm{TH}}$ is a predefined \textit{sensing radius}. The set of sensors that perform measurements at time $t$ is denoted by $G^s_t$. The sensing radius is chosen so as to ensure that a sufficient number of sensors can detect the target, but should be small enough that the model in \eqref{eq:measure-slat} remains valid.

\section{Simultaneous Sensor Localization and Target Tracking}\label{sec:slat} 

Our goal is to obtain the posterior marginal PMFs  (referred to as the \textit{beliefs}), $p({{\bf{x}}_t}|{{\bf{e}}_{1:t}})$ and $p({{\bf{z}}_{n,t}}|{{\bf{e}}_{1:t}})$, of the following joint distribution:
\beqa\label{eq:joint-pmf}
&p({{\bf{x}}_0},...,{{\bf{x}}_t},{{\bf{z}}_{1,0}},...,{{\bf{z}}_{{N_s},t}}|{{\bf{e}}_{1:t}})\propto&\\ 
&~~~~~~~ {p_0}({{\bf{x}}_0})\prod\limits_{n=1...N_s} {p_{n,0}}({{\bf{z}}_{n,0}})\prod\limits_{t' = 1...t\hfill\atop \forall n \in G_t^s} {p({d _{n,t}}|{{\bf{x}}_{t'}},{{\bf{z}}_{n,t'}})} \cdot&\nonumber\\
& \prod\limits_{t' = 1...t} {p(\mb{v}_{t'},{{\bf{x}}_{t'}}|{{\bf{x}}_{t' - 1}})} \prod\limits_{t' = 1...t\hfill\atop n = 1...{N_s}} {p({{\bf{z}}_{n,t'}}|{{\bf{z}}_{n,t' - 1}})},&\nonumber
\eeqa
where ${{\bf{e}}_{1:t}}$ is all available evidence up to time $t$ (i.e., measured TOAs and velocities). The previous factorization is obtained using Bayes' rule and standard assumptions \cite{Wymeersch2009}, such as independence of the measurements/priors and memoryless movement. Since the marginalization of (\ref{eq:joint-pmf}) is intractable, we resort to message-passing on a graphical model. 

\subsection{Graphical model}\label{subsec:graph-model}

We use an undirected graphical model \cite{Wainwright2008}, also known as a Markov random field (MRF),\footnote{An equivalent algorithm can be derived using the forward phase of the \textit{forward-backward} algorithm (also known as BCJR) in a hidden Markov model \cite{Rabiner1989}. However, we prefer to use a much more flexible framework, valid for discrete, continuous and mixed variables.} to represent the factorization in \eqref{eq:joint-pmf}. In a MRF, each vertex represents a random variable with an associated \textit{single-node} potential (a local evidence), and each edge represents a \textit{pairwise} potential (a likelihood function). An example is shown in Fig.~\ref{fig: slat-mrf}. Using the models defined in Sections~\ref{subsec:dyn} and \ref{subsec:measure}, the potentials are given by:

\begin{figure}[!t]
\centering
\includegraphics[width=0.95\columnwidth]{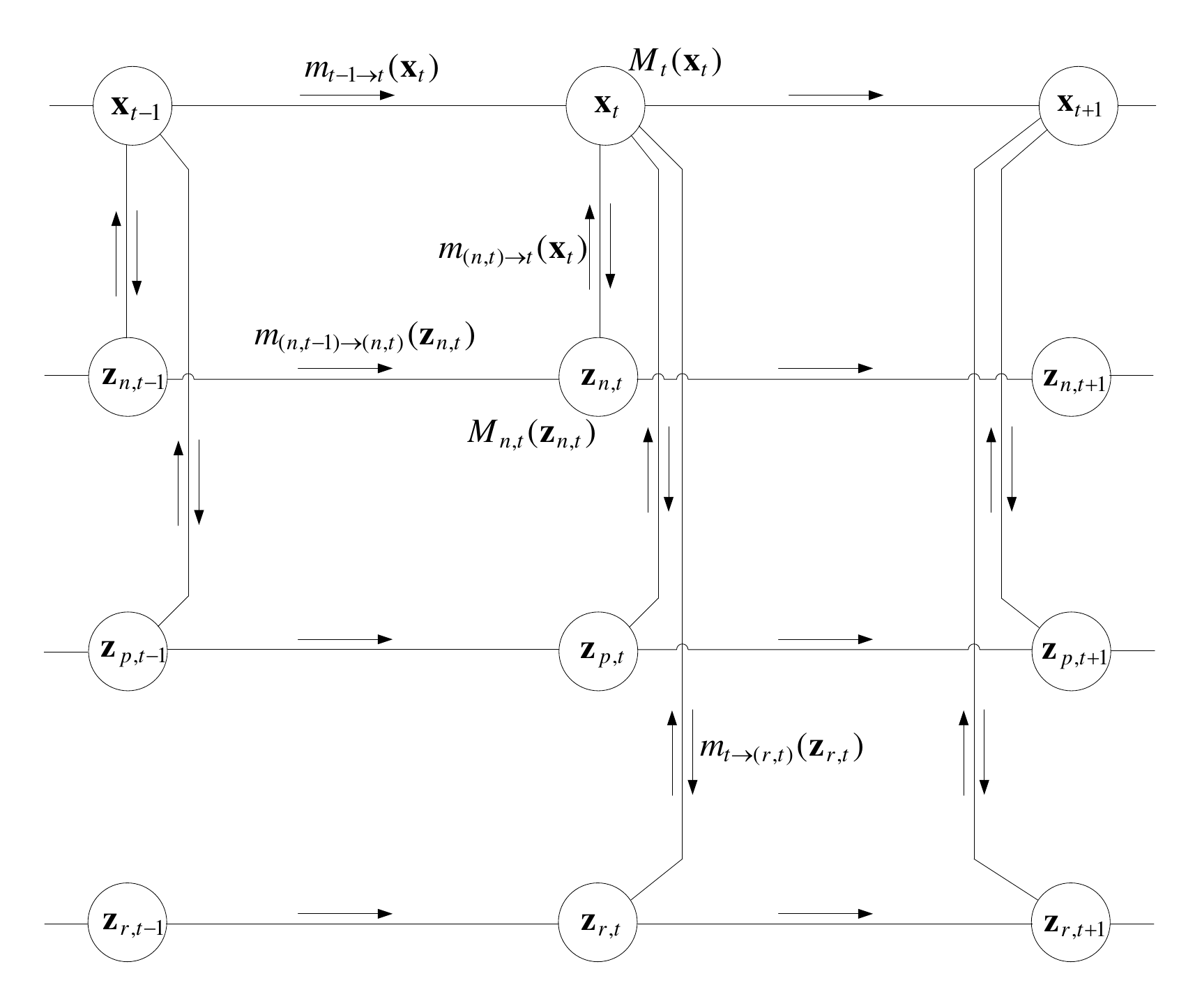}
\caption{An example of a MRF, with three sensors ($n$, $p$, and $r$) and a target at three time instants ($t-1$, $t$, and $t+1$). The messages are represented with arrows.} %
\label{fig: slat-mrf}
\end{figure}

\be\label{eq:single-slat1}
{\psi _t}({{\bf{x}}_t}) = \left\{ \begin{array}{l}
{p_0}({{\bf{x}}_0}),\,\,\,{\rm{if}}\,\,t = 0\\
1,\,\,\,\,\,\,\,\,\,\,\,{\rm{otherwise}}
\end{array}, \right.
\ee
\be\label{eq:single-slat2}
{\psi _{n,t}}({{\bf{z}}_{n,t}}) = \left\{ \begin{array}{l}
{p_{n,0}}({{\bf{z}}_{n,0}}),\,\,\,{\rm{if}}\,\,t = 0\\
1,\,\,\,\,\,\,\,\,\,\,\,{\rm{otherwise}}
\end{array}, \right.
\ee
\be\label{eq:pairwise-slat}
\psi_{t,(n,t)} (\mb{x}_t ,\mb{z}_{n,t}) =p(d_{n,t}|\mb{x}_t,\mb{z}_{n,t}),
\ee
\be\label{eq:pairwise-track}
\psi_{t-1,t} (\mb{x}_{t-1} ,\mb{x}_{t}) = p(\mb{v}_{t},{{\bf{x}}_t}|{{\bf{x}}_{t - 1}}),
\ee
\be\label{eq:pairwise-sens}
\psi_{(n,t-1),(n,t)} (\mb{z}_{n,t-1},\mb{z}_{n,t}) =p({{\bf{z}}_{n,t}}|{{\bf{z}}_{n,t - 1}}).
\ee 
Note that these potentials, as well as all messages and beliefs (defined in further text), are not necessarily normalized.

\subsection{SLAT via Real-Time Belief Propagation (BP)}\label{subsec:slat-bp} 

We adapt the standard BP (see \cite[eqs.~(8)-(9)]{Ihler2005a}) for our real-time and discrete problem. To ensure real-time execution, we do not send the messages backward in time (see also Fig. \ref{fig: slat-mrf}), and since the variables are discrete, we replace integration with summation. The beliefs (denoted with ${M_{n,t}}({{\bf{z}}_{n,t}})$ and ${{M}_t}({{\bf{x}}_t})$, respectively) are initialized with  ${M_{n,0}}({{\bf{z}}_{n,0}})={\psi _{n,0}}({{\bf{z}}_{n,0}})$ and  ${{M}_0}({{\bf{x}}_0})={\psi _0}({{\bf{x}}_0})$. The algorithm  (for time slot $t$, $t>0$) is summarized below:

\textbf{Step 1.} Compute the sensor-to-target and target-to-target messages:
\be\label{eq:msg-nt-slat}
{m_{(n,t) \to t}}({{\bf{x}}_t}) = \sum\limits_{{{\bf{z}}_{n,t}}} {{\psi _{t,(n,t)}}({{\bf{x}}_t},{{\bf{z}}_{n,t}}){M_{n,t - 1}}({{\bf{z}}_{n,t}})},
\ee
\be\label{eq:msg-time-slat}
{m_{t - 1 \to t}}({{\bf{x}}_t}) = \sum\limits_{{{\bf{x}}_{t - 1}}} {{\psi _{t - 1,t}}({{\bf{x}}_{t - 1}},{{\bf{x}}_t}){M_{t - 1}}({{\bf{x}}_{t - 1}})}.
\ee

\textbf{Step 2.} Update the target's belief:
\be\label{eq:belief-time-slat}
{{M}_t}({{\bf{x}}_t}) = {{m}_{t - 1 \to t}}({{\bf{x}}_t})\prod\limits_{n \in {G^s_t}} {{{m}_{(n,t) \to t}}({{\bf{x}}_t})}.
\ee

\textbf{Step 3.} Compute the target-to-sensor and sensor-to-sensor messages:
\be\label{eq:msg-tp-slat}
{m_{t \to (n,t)}}({{\bf{z}}_{n,t}}) = \sum\limits_{{{\bf{x}}_t}} {{\psi _{t,(n,t)}}({{\bf{x}}_t},{{\bf{z}}_{n,t}})\frac{{{M_t}({{\bf{x}}_t})}}{{{m_{(n,t) \to t}}({{\bf{x}}_t})}}},
\ee
\be\label{eq:msg-pp-slat}
{m_{(n,t - 1) \to (n,t)}}({{\bf{z}}_{n,t}}) = {M_{n,t - 1}}({{\bf{z}}_{n,t}}).
\ee

\textbf{Step 4.} Update the beliefs of the sensors:
\be\label{eq:belief-pt-slat}
\begin{array}{l}
{M_{n,t}}({{\bf{z}}_{n,t}}){\rm{ }} =  \left\{ \begin{array}{l}
{M_{n,t - 1}}({{\bf{z}}_{n,t}}){{m}_{t \to (n,t)}}({{\bf{z}}_{n,t}}),\,\,\,{\rm{if}}\,\,n \in G^s_t \\
{M_{n,t - 1}}({{\bf{z}}_{n,t}}),\,\,\,\,{\rm{otherwise}}
\end{array} \right.
\end{array}
\ee

\textbf{Step 5. (optional)} Compute the estimates using the k-nearest neighbour (kNN) approach \cite{Ni2003}:
\be\label{eq:estim-sens-slat}
{\hat{\bf{z}}_{n,t}} = \frac{\sum\limits_{{{\bf{z}}_{n,t}} \in \,C_{{{\bf{z}}_{n,t}}}^k} {{{\bf{z}}_{n,t}}{M_{n,t}}({{\bf{z}}_{n,t}})}}{\sum\limits_{{{\bf{z}}_{n,t}} \in \,C_{{{\bf{z}}_{n,t}}}^k} {{M_{n,t}}({{\bf{z}}_{n,t}})}},
\ee
\be\label{eq:estim-track-slat}
{\hat{\bf{x}}_t} = \frac{\sum\limits_{{{\bf{x}}_t} \in \,C_{{{\bf{x}}_t}}^k} {{{\bf{x}}_t}{M_t}({{\bf{x}}_t})}}{\sum\limits_{{{\bf{x}}_t} \in \,C_{{{\bf{x}}_t}}^k} {{M_t}({{\bf{x}}_t})}},
\ee
where $C_{{{\bf{z}}_{n,t}}}^k$ and $C_{{{\bf{x}}_t}}^k$ are the set of $k$ cells with highest beliefs, $M_{n,t}({{\bf{z}}_{n,t}})$ and ${M_t}({{\bf{x}}_t})$, respectively. The special cases $k=1$ and $k=N_c$, correspond to MAP and MMSE estimates, respectively. Note that this phase is optional, since the main output of this algorithm (beliefs) have been already computed in steps 2 and 4.

For comparison purposes, we also consider two specific instances of the proposed SLAT algorithm: i) Bayesian point-mass target tracking (by excluding all target-to-sensor messages, i.e., ${m_{t \to (n,t)}}({{\bf{z}}_{n,t}})=1$), and ii) Bayesian point-mass target localization (by excluding all target-to-sensor and target-to-target messages, i.e., ${m_{t \to (n,t)}}({{\bf{z}}_{n,t}})=1$ and  ${m_{t - 1 \to t}}({{\bf{x}}_t})=1$). The former one uses the sensors' priors to track the target, while the latter one uses the sensors' priors to locate the target independently in each time slot (i.e., without a dynamic model). Note that the target tracking algorithm in \cite{Hedley2013} can be considered as a special case of SLAT-BP (although their measurement and dynamic models are different).

\subsection{Implementation Issues}\label{subsec:issues} 

In this section, we discuss some important issues that can arise during the implementation of the proposed SLAT-BP algorithm.
\bi
\item \textit{Complexity of the algorithm:} The complexity of the SLAT-BP algorithm at time $t$ is $\mathcal{O}(\left|G^s_t\right|N_c^2)$, since the  message computations dominates the cost. This complexity is significantly  less than that of naive marginalization of (\ref{eq:joint-pmf}) which would require $\mathcal{O}(N_c^{N_s+t-1})$ operations at time instant $t$. Although this is a significant reduction, the complexity is still high if there are many cells. The complexity can be further reduced by considering only beliefs with a probability larger than a predefined \textit{belief threshold} $\epsilon_M$ ($\epsilon_M<1$). For example, in \eqref{eq:msg-nt-slat}, only cells $c$, which satisfy the following constraint:
\be\label{eq:bthreshold}
{M_{n,t - 1}}(\mb{l}_c)/\sum\limits_{c'=1\ldots N_c}{M_{n,t - 1}}(\mb{l}_{c'})>\epsilon_M/N_c,
\ee
should be considered for summation. An analog constraint is then used for \eqref{eq:msg-time-slat} and \eqref{eq:msg-tp-slat}. Denoting the number of these cells by $N^{\epsilon}_{{\bf{z}}_{n,t}}$ and $N^{\epsilon}_{{\bf{x}}_{t}}$, and $N^{\epsilon}_t=\max(\max_{n \in G^s_t}(N^{\epsilon}_{{\bf{z}}_{n,t}}),N^{\epsilon}_{{\bf{x}}_{t}})$, the complexity at time $t$ is reduced to $\mathcal{O}(\left|G^s_t\right|N_c N^{\epsilon}_t)$.

\item \textit{Non-synchronized measurements:} In Section~\ref{subsec:dyn}, we assumed that the IMU is configured to operate at the same rate as the sensors (reporting every $T_s$ second). In practice, the rate of the IMU may be much higher than that of the sensors. The proposed message-passing algorithm can be easily adapted to a situation where the rates are different. Assuming that the algorithm operates at the IMU rate and that we do not know the rate of the sensors' measurements, we just need to do following at each time slot: i) if the sensors' measurements are available, we run the   algorithm in Section~\ref{subsec:slat-bp}, and  ii) if the sensors' measurements are unavailable,   simply exclude all sensor-to-target messages (i.e., ${m_{(n,t) \to t}}({{\bf{x}}_t})=1$). In other words, we run simultaneous localization and dead reckoning in all time slots in which only the IMU measurements are available.

\item \textit{Routing data to the FCs:} All collected measurements should be routed to the FCs as soon as they become available. Although there are many well-known routing protocols \cite{Royer1999}, we recommend a hybrid system based on a \textit{leaky-feeder system} (LFS) and a \textit{wireless mesh network} (WMN), similar to one installed at a coal mine in West Virginia, US \cite{Novak2009}. LFS consists of a coaxial-type cable, which emits and receives radio waves (i.e, it behaves as a distributed antenna). It has many power supplies, and a backup battery in an explosion-proof enclosure. Therefore, all data transmitted by the sensors or the target, will be available to FCs using a one-hop communication link and without any routing protocol. Since LFS typically cannot provide coverage all over the deployment area, it should be complemented with a WMN, which should consist of the subset of WSN not in the vicinity of the LFS cables. In the WMN, sensors communicate in a multi-hop fashion using an optimal path, computed in real-time. If one or a few sensors fails, the system simply recomputes the path. Therefore, this system is capable of routing the data as long as there is one path between a sensor and a FC or LFS cable. 

\item \textit{Online calibration:} Although the measurement model in \eqref{eq:measure-slat} provides  robustness against dynamic obstacles, it will not be good enough if a permanent change is made to the environment (e.g., new pillars are formed, or the tunnels are extended). In that case, it would be necessary to repeat the calibration (especially, to re-estimate the GM parameters for the NLOS error model), which is a cumbersome task. An alternative, and preferable option, is to update  the measurement model online using already deployed sensors. This is feasible since the fusion center knows the current estimates of the sensors' positions, and consequently all inter-sensor distances. The additional requirement is that the sensors are randomly deployed so that they can provide sufficient statistics for parameter estimation. This calibration should be done periodically (e.g., once per day), or manually triggered once some change in the deployment area is reported.
\ei

\begin{figure*}[!t]
\centerline{
\subfloat[]{\includegraphics[width=0.75\textwidth]{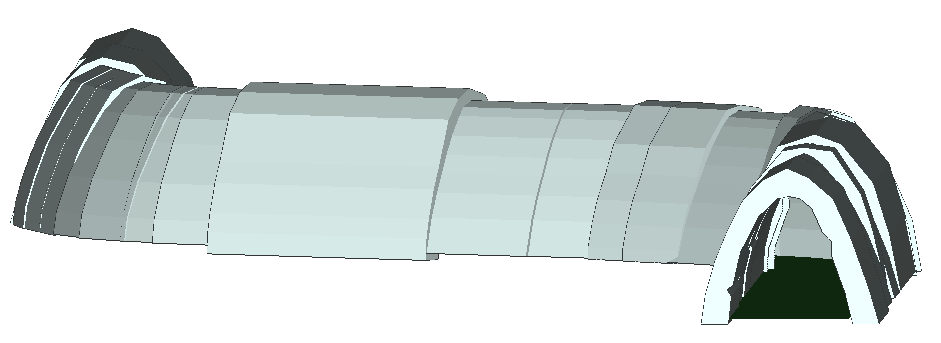}\label{fig:Tunnel1-3D}}
}
\centerline{
\subfloat[]{\includegraphics[width=0.75\textwidth]{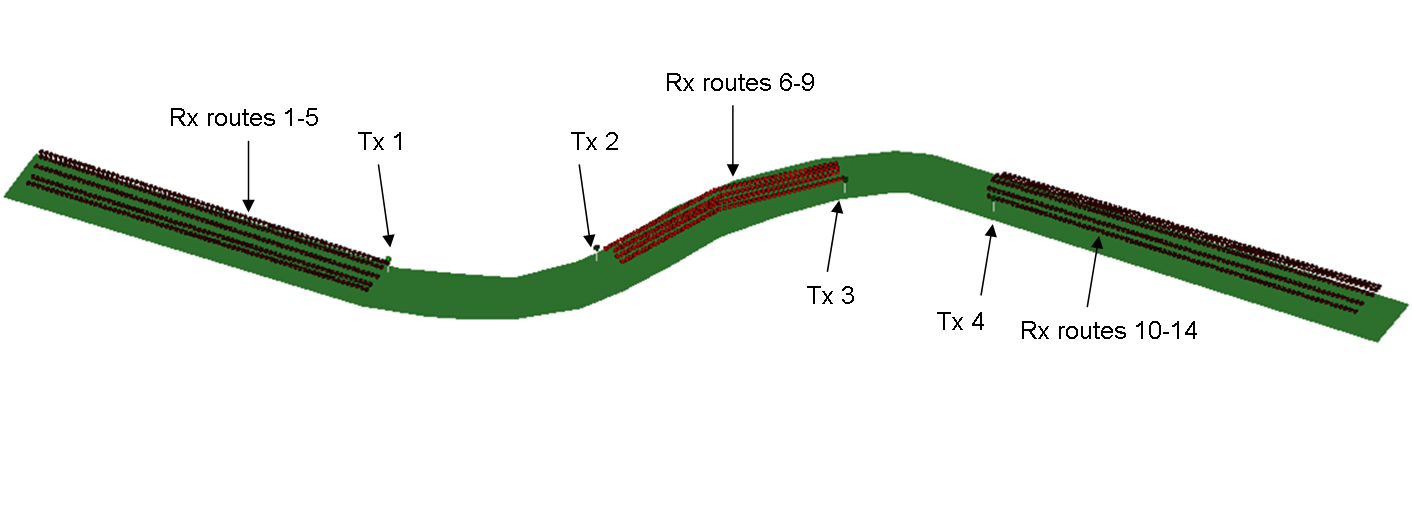}\label{fig:deploymentNLOS}}
}
\caption{Illustrations of the designed tunnel in Wireless InSite: (a) 3D illustration, and (b) deployment of Tx/Rx through the tunnel. The dimensions of the tunnel are approx. 110m (length) x 5m (width) x 5m (height) (corresponding to example in Fig. \ref{fig:mine-cells}a).}
\label{fig:tunnel-txrx}
\end{figure*}

\section{Numerical results}\label{sec:sims}

In this section, we analyze the accuracy and the robustness of the proposed approach using ray-tracing simulation.

\subsection{TOA error modelling using ray-tracing}\label{subsec:ray-tracing}

For TOA error modelling, we decided to use REMCOM's Wireless InSite ray-tracing simulator \cite{REMCOM,REMCOM2012manual}. Wireless InSite is a flexible, powerful tool for accurately predicting the effects of the environment on the propagation of electromagnetic waves. It models the physical
characteristics of the environment (including the effects of terrain, urban buildings and tunnel features), performs the electromagnetic calculations, and evaluates the signal propagation characteristics. The calculations are made by shooting rays from the transmitters, and propagating them through the environment, until they arrive at the receivers. The rays interact with environmental features via: \textit{reflections} at object faces, \textit{diffraction} around objects, and \textit{transmission} (penetration) through objects. Wireless InSite can provide quantities such as electric and magnetic field strength, received signal strength, time of arrival, path-loss, delay spread, direction of arrival, impulse response and power delay profile. 

We designed an artificial mine tunnel using Wireless InSite (see Fig.  \ref{fig:Tunnel1-3D}), by creating many small pieces, and connecting them together. The dimensions of the tunnel are similar to a real mine tunnel \cite{Nerguizian2006}, and each piece contains expected irregularities. We chose concrete as material for the tunnel walls, which corresponds to the areas in the mine reinforced to increase stability. Then, we deployed transmitters (Tx) and receivers (Rx) routes along the tunnel, as shown in Fig.~\ref{fig:deploymentNLOS}. Our goal is to obtain TOA at each receiver, which has NLOS to the transmitter caused by the tunnel walls. LOS links are not considered since they provide ground-true estimates (the TOA estimates from Wireless InSite do not include other sources of errors, such as thermal noise, or limited bandwidth). Finally, we set the parameters\footnote{Many other parameters (see \cite{REMCOM2012manual}), not shown in Table \ref{table:param}, are kept to default values since they are irrelevant for this analysis.} as shown in Table \ref{table:param}. We chose the Full-3D Shooting and Bouncing Ray (SBR) method, which includes the effect of reflections, transmissions, and diffractions on the electric field in 3D environment, without any restriction on the object shapes (more details about this method, and possible alternatives can be found in \cite[Chapter 15]{REMCOM2012manual}). Regarding maximum number of reflections, transmissions, and diffractions, taking into account the analyses in \cite{Chuasomboon2013}, we chose the values which provide a good trade-off between the performance and complexity (i.e., any further increase of these values would lead to a negligible difference in the results, while the computational time would grow dramatically). We used short-dipole antennas with vertical polarization in order to ensure a near-omni-directional radiation pattern. Finally, the values for input power, transmission line loss, and received power threshold were chosen to ensure sufficient range for ray propagation.

\begin{table}[!tb]
\caption{Main parameters used for Wireless InSite simulations}
\label{table:param}
\centering
\begin{tabular}{l l}
\hline 
Ray-tracing method & Full-3D SBR\\
Antenna (Tx/Rx) & short dipole\\
Polarization (Tx/Rx) & vertical\\
Relative permittivity of the wall & 7\\
Conductivity of the wall & 0.015 $\mathrm{m}^{-1}\Omega^{-1}$\\
Waveform (Tx/Rx) & sinusoidal\\
Central frequency & 2.4 GHz\\
Input power (Tx) & 15 dBm\\
Received Power Threshold (Rx) & -110 dBm\\
Transmission line loss (Tx/Rx) & 0 dB\\
Altitude of antenna (Tx/Rx) & 1.3 m\\
Maximum number of reflections & 8\\
Maximum number of transmissions & 1\\
Maximum number of diffractions & 1\\
\hline 
\end{tabular}
\end{table}

\begin{figure}[!t]
\centering
\includegraphics[width=0.95\columnwidth]{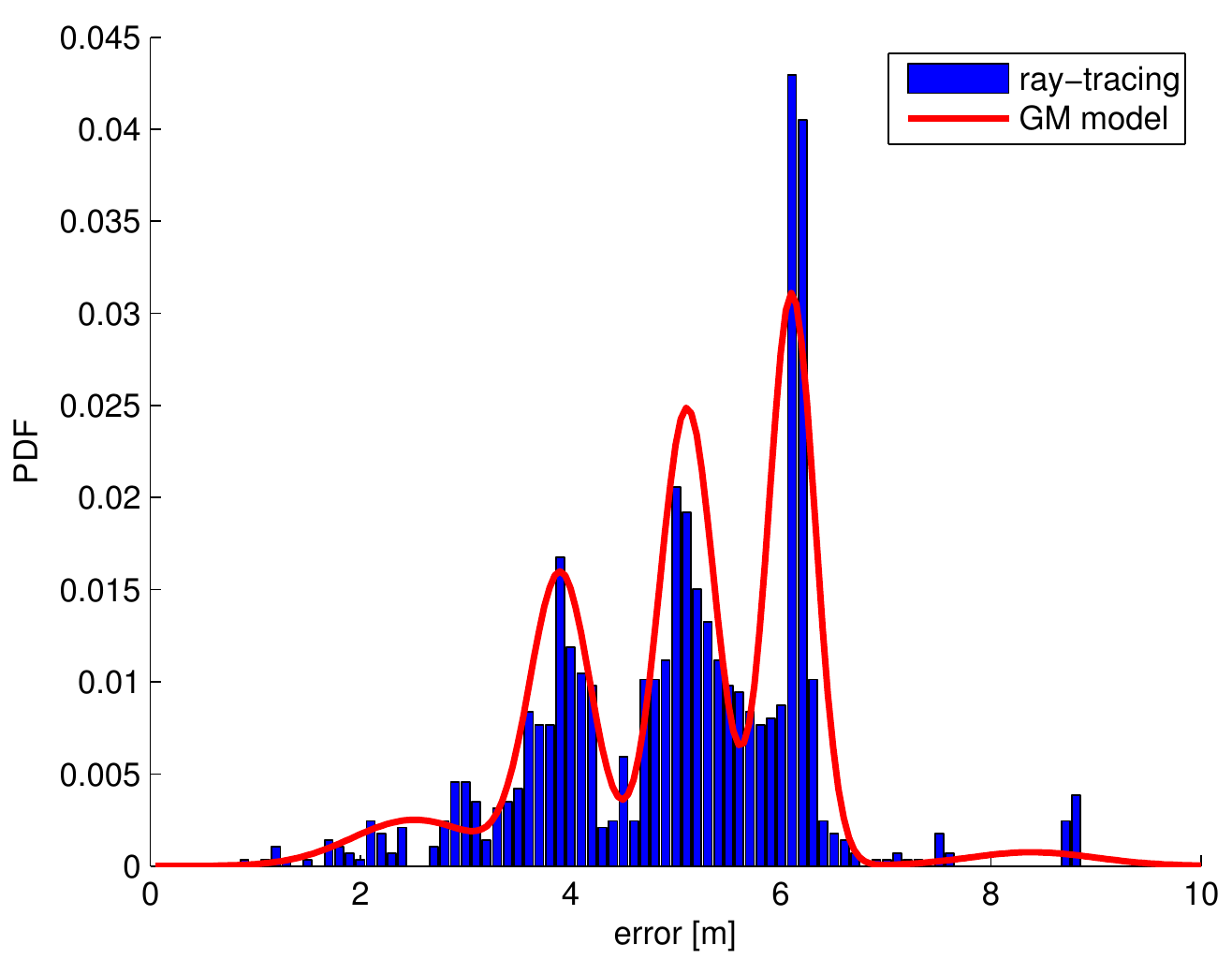}
\caption{Histogram of the NLOS ranging error, and corresponding Gaussian mixture model.}
\label{fig:pdf-gmm-nlos}
\end{figure}

From  the simulation, we collected 1130 impulse responses, and obtained the same number of TOA samples (in the absence of noise, the threshold for TOA estimation is set to zero). The samples were used to find the distance estimates, and the samples of the ranging error (the difference between the true and the estimated distance), in order to create a GM model. Observing the histogram in Fig.~\ref{fig:pdf-gmm-nlos}, we can see that the error is positive (as expected in the absence of the other noise), and that it can be well approximated with 5 modes. Moreover, we can see that the error is not exponentially distributed, which stands in contrast to results \cite{Chen1999,Venkatesh2007} for other environments. Therefore, we set the number of GM components to $N_M=5$, and run the k-means algorithm to obtain its parameters.\footnote{The obtained parameters are not provided since they are valid only for this, artificially generated, tunnel. For other environments, the error modelling should be repeated.}

According to Fig. \ref{fig:pdf-gmm-nlos}, this model provides a good (but not perfect) approximation of the real PDF. We did not  include other error sources  in the NLOS model, since they typically are negligible compared to the multipath error (recall that the signal-to-noise-ratio is very high for short-range sensing, and that we target WB/UWB signals). For the LOS error model, we did not analyze the precise impact of the transmit power, bandwidth and other sources of the error. We simply assumed that it can be well approximated with a zero-mean Gaussian distribution with a sufficiently large standard deviation $\sigma_{w,0}= c \cdot 3.33~\mathrm{ns}= 1$ m.

Finally, we considered a scenario with the same deployment as in
Fig.~\ref{fig:deploymentNLOS}, but with different types of the
obstacles (objects created of metal, water, sand, etc.) placed in
front of the transmitters. We found that the bias caused by these
obstacles is very uncertain, so our assumption that it is uniformly
distributed (as a least informative option) is reasonable. Assuming
that these obstacles will block the LOS path in $3\%$ of the cases, we
increased the size of the NLOS database to 1164 samples by adding
(randomly chosen) samples from the NLOS scenarios with obstacles. This
database, along with the GM model, will be used in the next sections
as an input to the performance analysis simulations.

\subsection{Simulation Setup}\label{subsec:sim-setup}

We considered the tunnel in Fig.~\ref{fig:Tunnel1-3D}
divided in $N_c=44$ cells, and with $N_s=25$ sensors randomly
deployed in these cells. The sensors' priors were
given by ${p_{n,0}}({{\bf{z}}_{n,0}}) = {\cal
  N}({{\bf{z}}_{n,0}};{{\bf{l}}_n},{\Sigma _S})$ where ${{\bf{l}}_n}$
($n=1,\ldots,N_s$) is the reported location provided by the deployment
team, and ${\Sigma _S} =
{\rm{diag}}(\sigma_S^2,\,\sigma_S^2,\,\sigma_S^2\,)$ ($\sigma_S=6$ m)
is the empirical measure of the precision of the sensor placement. We
assume that we know the initial cell of the target, i.e.,
$p_{0}({\bf{x}}_{0})=\delta({\bf{x}}_{0}-{{\bf{l}}_1})$. There are
$N_T=40$ time slots, and the sampling interval is $T_s=1$ s. The
target position at time $t$ is generated using the following mobility
model: \be\label{eq:mob-model} {{\bf{x}}_t} = \left\{ \begin{array}{l}
    {{\bf{l}}_{2t + \eta }}\,,\,\,\,\,\,\,\,\,\,\,\,\,\,\,\,\,\,\,\,\,\,\,{\rm{for}}\,\,1 \le t \le N_T/2+1,\\
    {{\bf{l}}_{2({N_c} - 1 - t) + \eta }}\,,\,\,{\rm{for}}\,\,N_T/2+2
    \le t \le {N_T}
\end{array}, \right.
\ee
where $\eta$ is a random integer between $-1$ and $1$ (i.e., $\eta \sim \rm{Unif}\{-1,0,1\}$), which adds uncertainty to the mobility of the target. This model assumes that the target is moving forward through the tunnel during the first half of the time, and then going backward until the end of the period. Recall that this model is not known to our algorithm, but the target's velocity is measured by the IMU.

We set the remaining parameters for the measurement noise as follows:
$P_{\mathrm{NLOS}}=0.17$, $P_{\mathrm{OBS}}=0.03$, and $\sigma_u=0.5$
m/s. The measurements are generated using these models, except in the
NLOS case, in which we pick a randomly-chosen sample from the
ray-tracing database. To take into account the cell size, we set the
quantization noise parameter to $D=5$ m. Moreover, the sensing radius was set to
$d_{\mathrm{TH}}=30$ m, the maximum distance error to
$D_{\mathrm{max}}=d_{\mathrm{TH}}=30$ m, the parameter for kNN
estimation to $k=2$, and the belief threshold to
$\epsilon_M=0.05$. All results were averaged over at least
$N_{\mathrm{MC}}=100$ Monte Carlo runs. In each run, we generated
different observations, target tracks and sensor positions.

All parameters, summarized in Table \ref{table:all-param}, were used unless otherwise stated in the following text.

\begin{table}[!tb]
\caption{Summary of parameters used in simulations}
\label{table:all-param}
\centering
\begin{tabular}{l l}
\hline
Parameter & Value\\
\hline
number of cells ($N_c$)&  44\\
number of sensors ($N_s$)&  25\\
number of time slots ($N_T$)&  40\\
sampling interval  ($T_s$)&  1 s\\
std. deviation of sensors' positions ($\sigma_S$) &  6 m\\
prob. of NLOS (tunnel wall)  ($P_{\mathrm{NLOS}}$)&  0.17\\
prob. of NLOS (obstacle)  ($P_{\mathrm{OBS}}$)&  0.03\\
std. deviation of IMU noise  ($\sigma_u$)&  0.5 m/s\\
std. deviation of LOS noise  ($\sigma_{w,0}$)&  1 m\\
sensing radius ($d_{\mathrm{TH}}$) &  30 m\\
maximum distance error ($D_{\mathrm{max}}$)&  $d_{\mathrm{TH}}$\\
quantization noise ($D$)&  5 m\\
kNN parameter ($k$) &  2\\
belief threshold ($\epsilon_M$) & $0.05$\\
number of Monte Carlo runs ($N_{\mathrm{MC}}$) & 100\\
\hline
\end{tabular}
\end{table}

\subsection{Performance Analysis}\label{subsec:perform}

Our goal is to analyze the accuracy of the proposed SLAT and compare
it with corresponding tracking and localization algorithms (defined in
Section~\ref{subsec:slat-bp}). The target's (sensor's) position error
is defined as the Euclidean distance between the position of the true
and the estimated cell of the target (sensor).

\begin{figure*}[!t]
\centerline{
\subfloat[]{\includegraphics[width=0.45\textwidth]{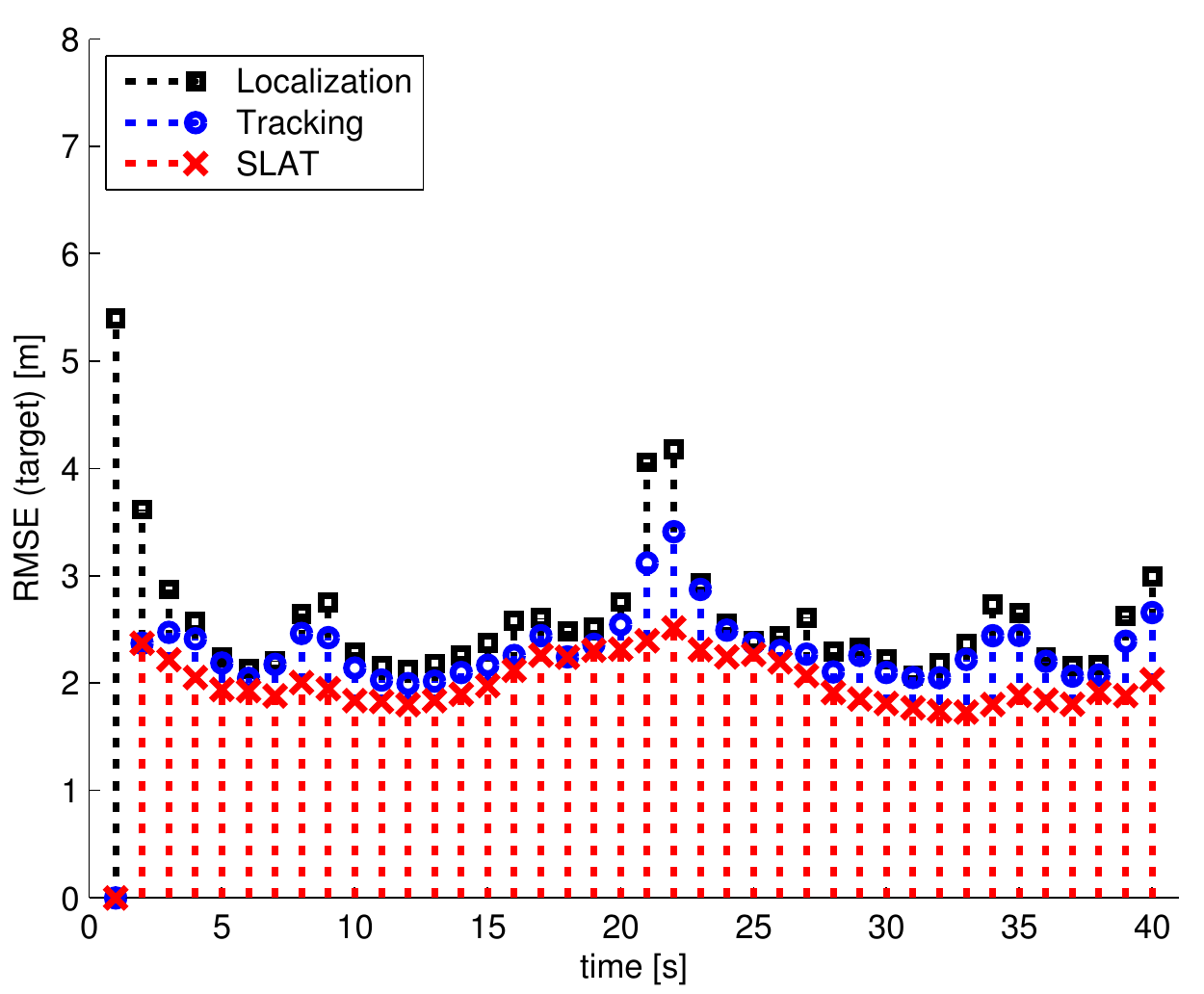}\label{fig:rmse-time-targ}} 
\hfil
\subfloat[]{\includegraphics[width=0.45\textwidth]{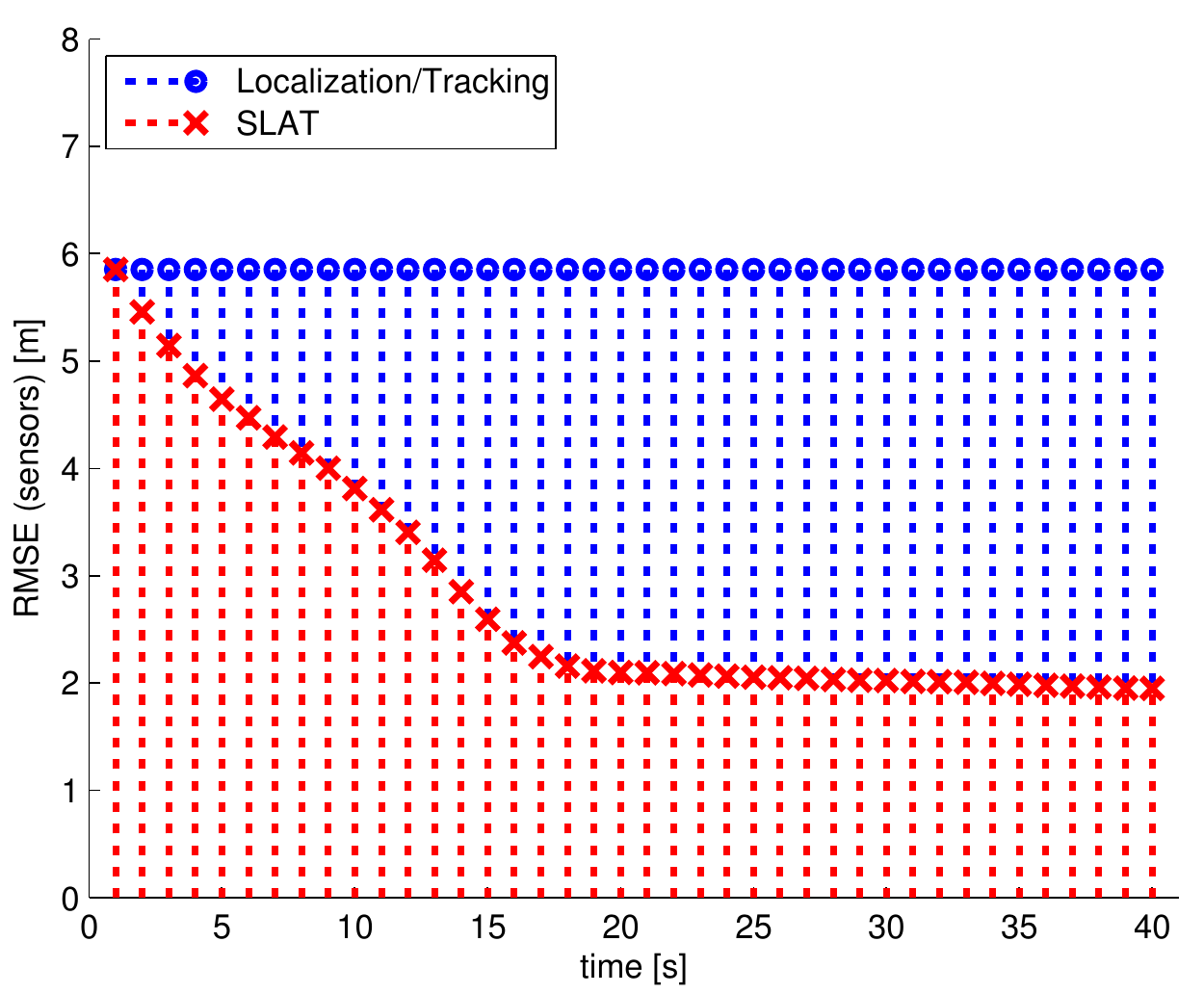}\label{fig:rmse-time-sens}}
}
\caption{RMSE as a function of time for (a) the target's position error, and (b) the sensors' position error. For the sensors, there is an additional averaging over all sensors' position errors.}
\label{fig:rmse-time}
\end{figure*}

\begin{figure*}[!ht]
\centerline{
\subfloat[]{\includegraphics[width=0.45\textwidth]{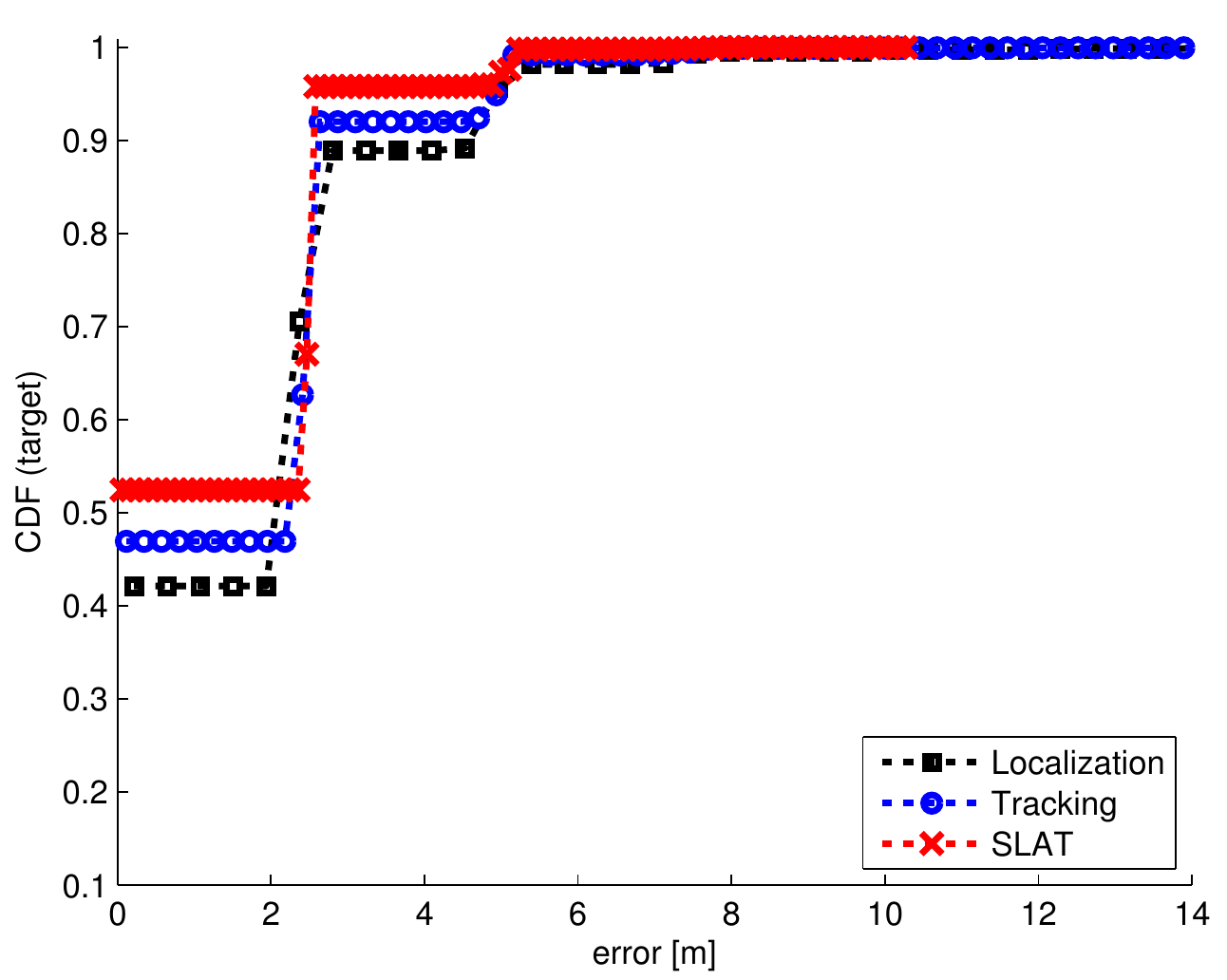}\label{fig:cdf-targ}}
\hfil
\subfloat[]{\includegraphics[width=0.45\textwidth]{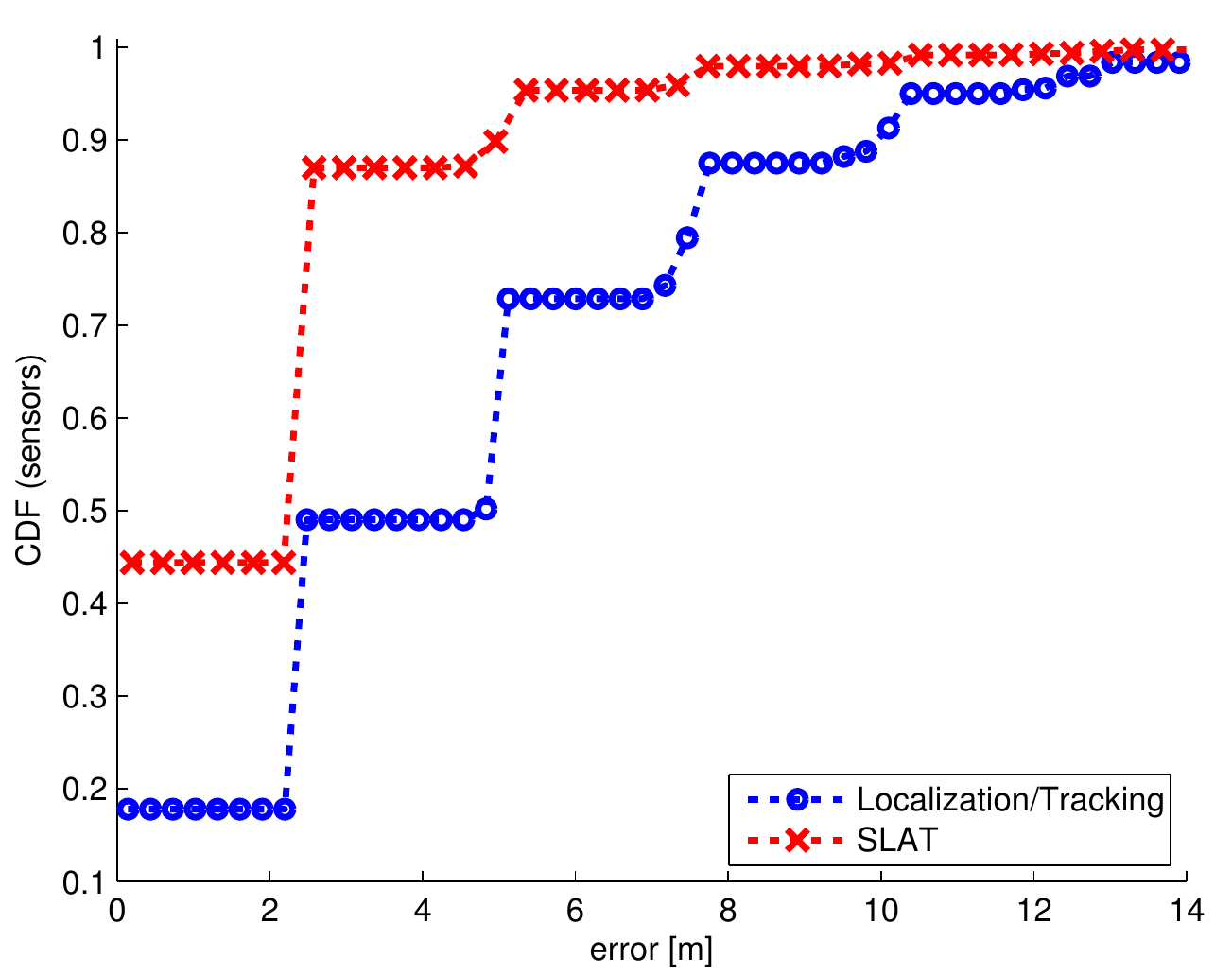}\label{fig:cdf-sens}}
}
\caption{CDF of (a) the target's position error, and (b) the sensors' position error.}
\label{fig:cdf-error}
\end{figure*}

\begin{figure*}[!t]
\centerline{
\subfloat[]{\includegraphics[width=0.33\textwidth]{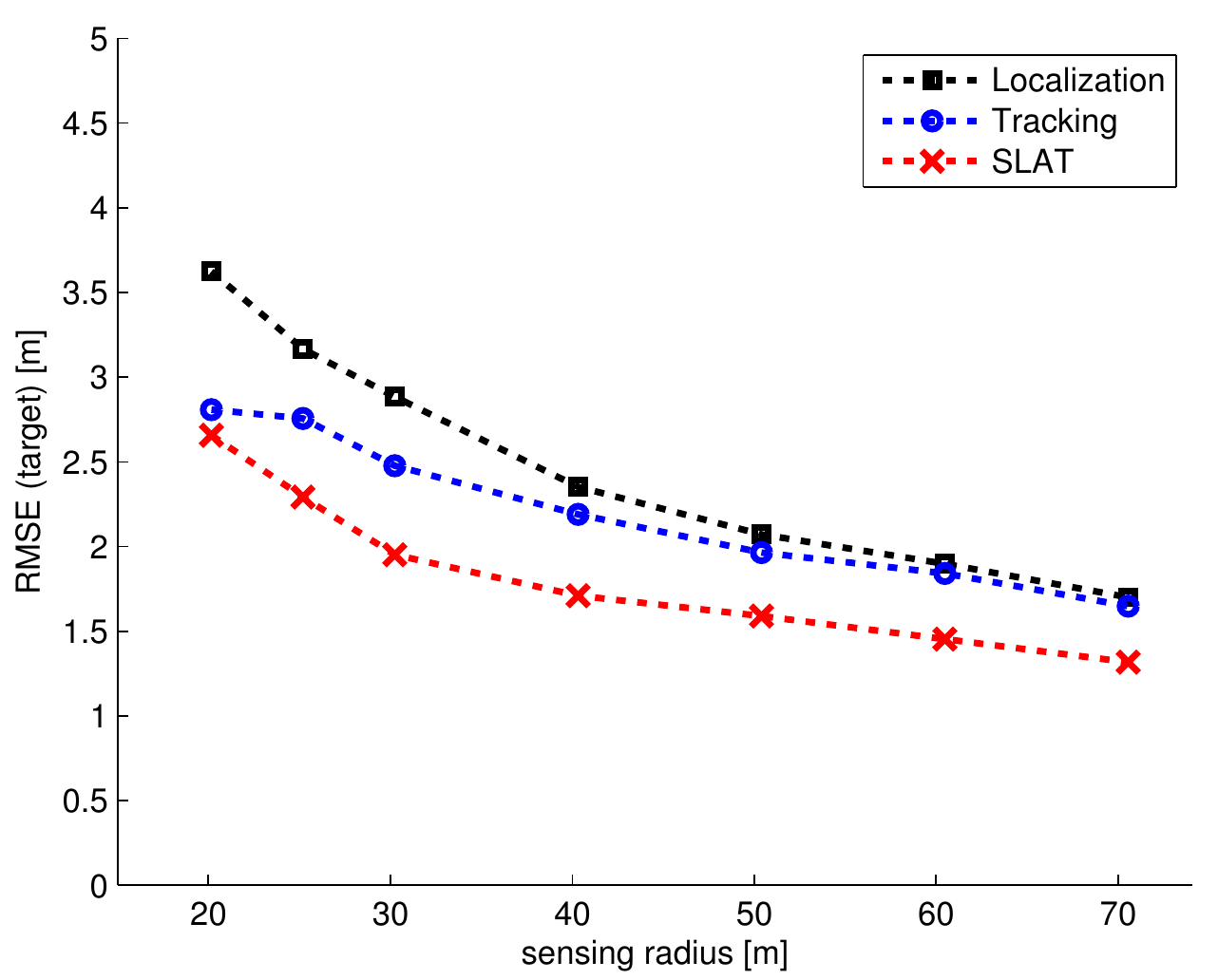}}\label{fig:rmse-thre-targ}
\hfill
\subfloat[]{\includegraphics[width=0.33\textwidth]{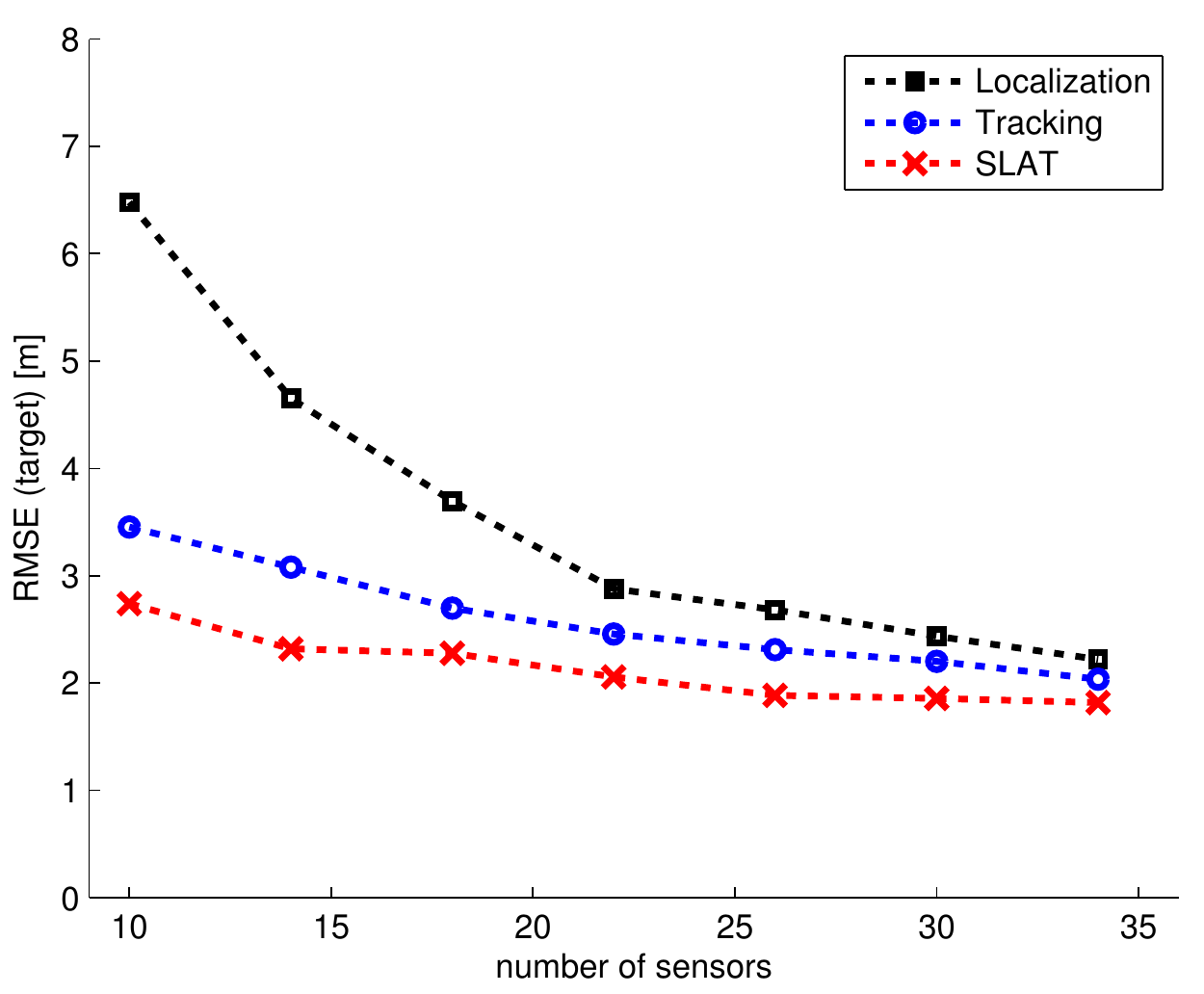}}\label{fig:rmse-node-targ}
\hfill
\subfloat[]{\includegraphics[width=0.33\textwidth]{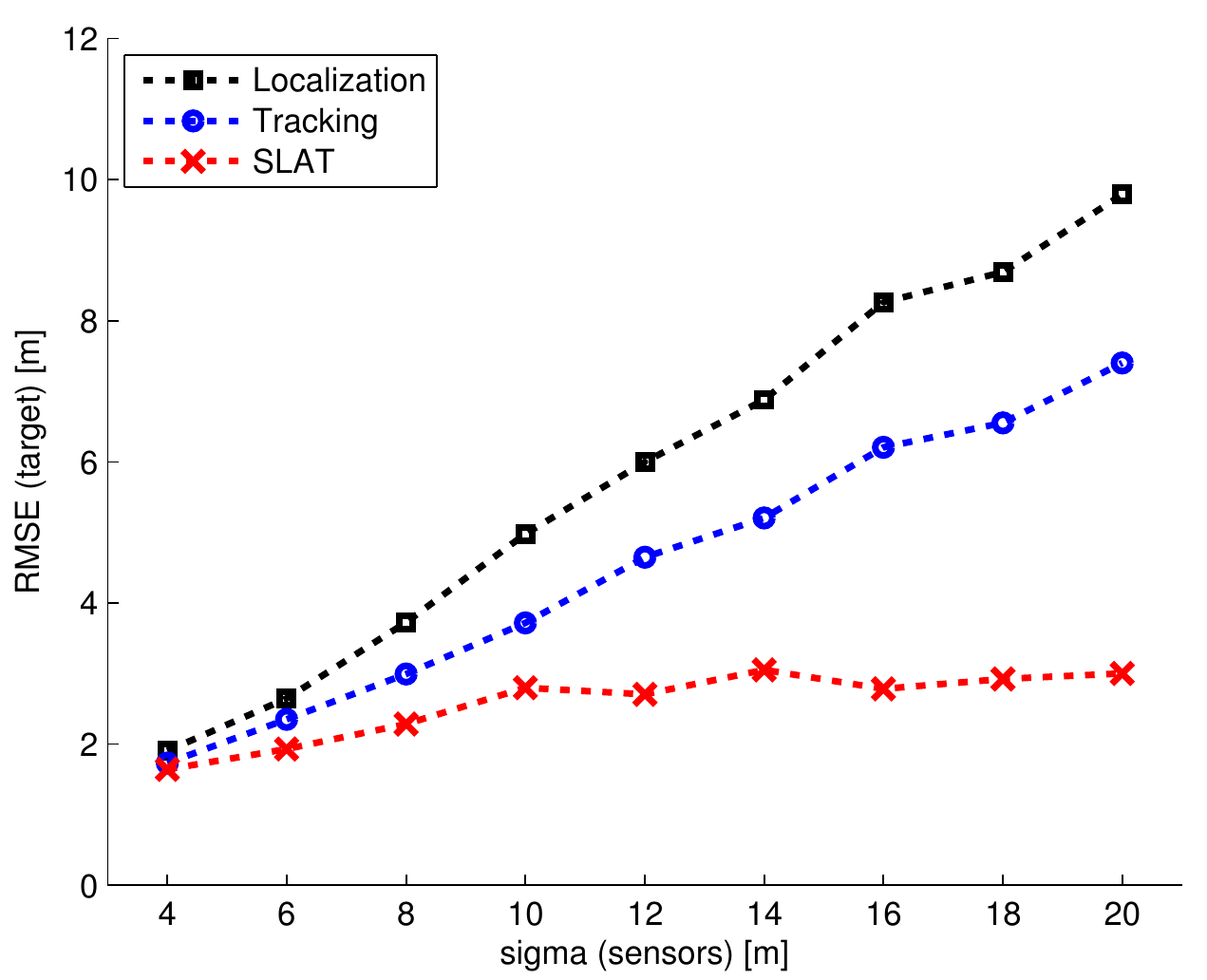}}\label{fig:rmse-sigm-targ}
}
\centerline{
\subfloat[]{\includegraphics[width=0.33\textwidth]{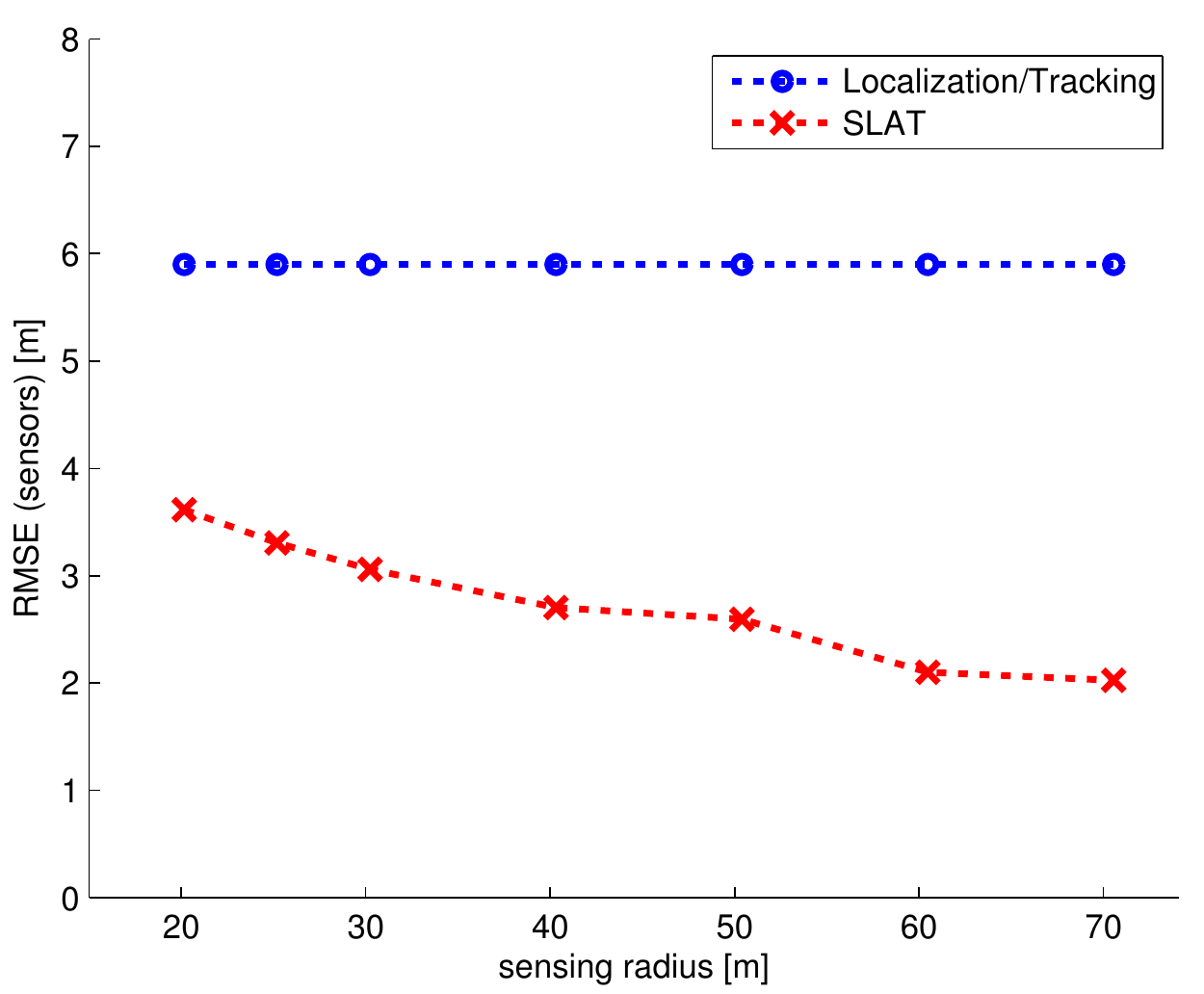}}\label{fig:rmse-thre-sens}
\hfill
\subfloat[]{\includegraphics[width=0.33\textwidth]{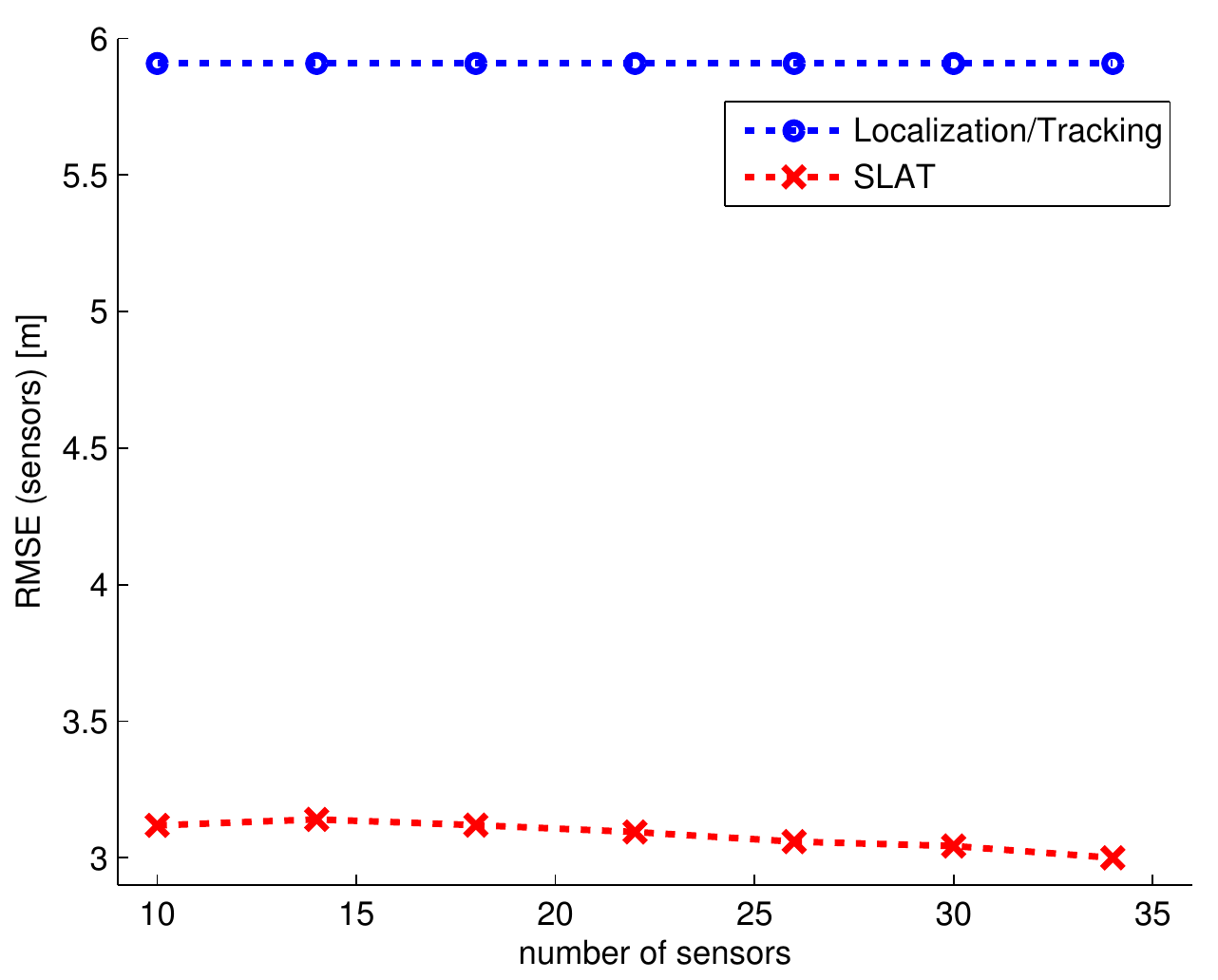}}\label{fig:rmse-node-sens}
\hfill
\subfloat[]{\includegraphics[width=0.33\textwidth]{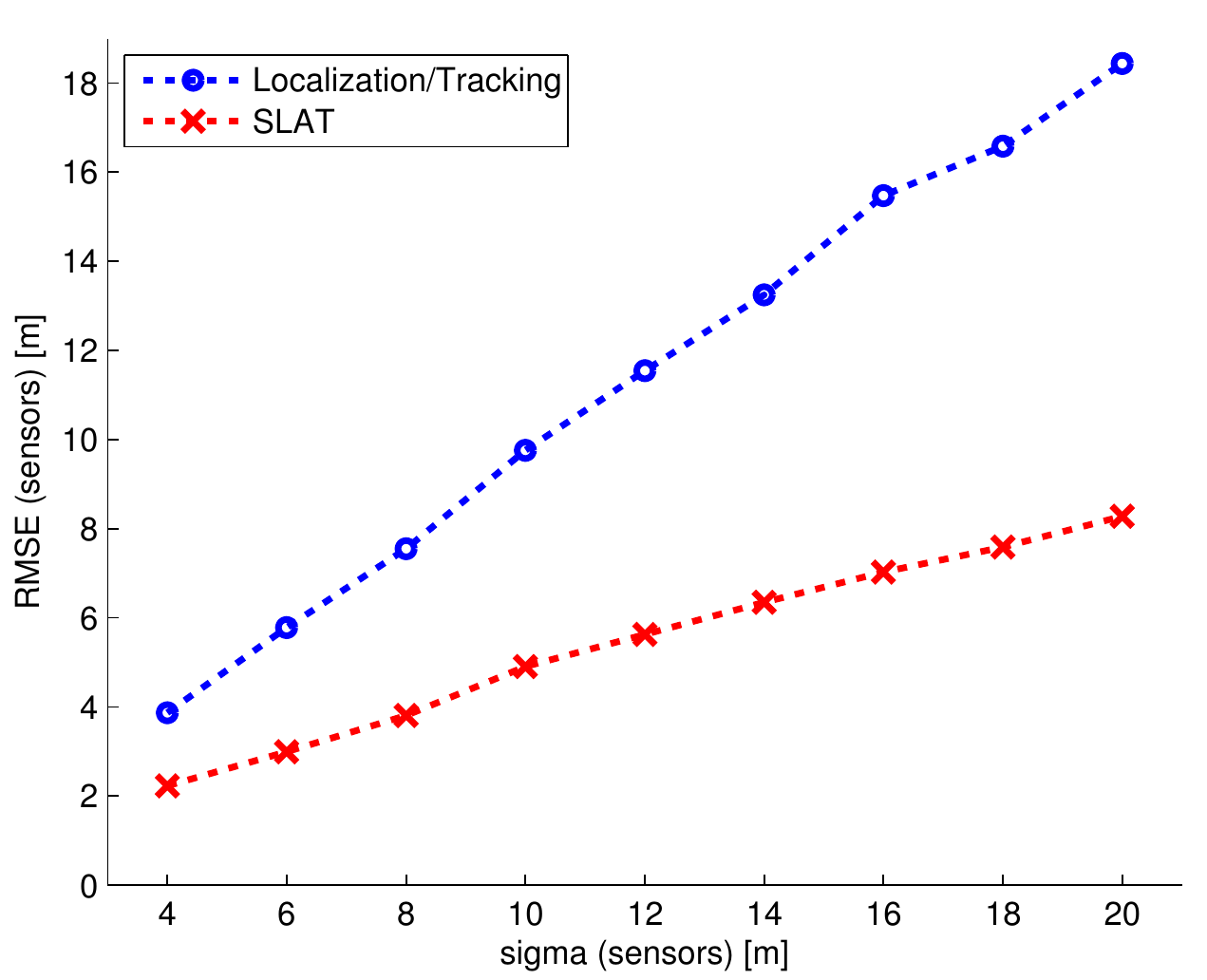}}\label{fig:rmse-sigm-sens}
}
\caption{RMSE as a function of: (a), (d) sensing radius, (b), (e) number of sensors, and (c), (f) standard deviation of the sensors position. Plots with RMSE of the target's position error are shown in the first row, while plots with RMSE of the sensors' positions error are shown in the second row.}
\label{fig:rmse-param}
\end{figure*}

We first analyze the root-mean-square error (RMSE) as a function of
time, as shown in Fig. \ref{fig:rmse-time}. Regarding the estimates of
the target positions (Fig. \ref{fig:rmse-time-targ}), SLAT provides
the best performance, followed by tracking, and localization which
provides the worst performance. The difference between tracking and
localization is caused by information from the IMU. The SLAT performs
better than tracking because the target uses the information from
improved sensors' PMFs, while only the information from initial
sensors' PMFs in the tracking algorithm. We also note that the
performance of all algorithms is the worst when the target is close to
the tunnel edges ($t=1,21,22,40$). This behavior is caused by the
small number of the sensors in proximity of the target. Regarding the
estimates of the sensors' positions (Fig. \ref{fig:rmse-time-sens}),
the SLAT consistently improves their estimates comparing with
localization/tracking which do not update the sensors' PMFs. This can
be explained by the fact that the state of the same variable (in
contrast to tracking) is estimated in each time slot. However, the
RMSE will not converge to zero because the NLOS measurement model is
not fully consistent with the measurements (see
Fig. \ref{fig:pdf-gmm-nlos}).

To draw more precise conclusions, we analyze the cumulative distribution
function (CDF)\footnote{The CDF contains discrete steps due to the
 finite number of possible error values.} of the target's and the
sensors' position error in Fig. \ref{fig:cdf-error}.  We can see in
Fig.~\ref{fig:cdf-targ} that the localization, tracking and SLAT
algorithms correctly estimate the target's cell in about 42\%, 46\%,
and 53\% of the considered tests, respectively. Regarding the sensor
position estimates (Fig.~\ref{fig:cdf-sens}), the
localization/tracking algorithms correctly estimate the sensors' cells
in only 18\% of the tests, while SLAT does so in about 45\% of the
tests. More importantly, the 95th percentile error of the SLAT (for
both the sensors' and target's RMSE) is at most the half of the
corresponding localization/tracking algorithms. Therefore, we can
conclude that SLAT outperforms tracking/localization in terms of both
the target's and the sensors' error. In both cases, SLAT is
consistently (at any percentile), better than the other two
algorithms. It is worth noting that these conclusions are consistent
with our previous results \cite{Savic2013fusion} based on models from
the CANMET mine.

We now analyze the effect of the different parameters on the
performance. The results are shown in Fig. \ref{fig:rmse-param}. We make the following observations: 
\bi
\item \textit{Effect of varying the sensing radius
    (Figs.~\ref{fig:rmse-param}a and \ref{fig:rmse-param}d):}
  Increasing the sensing radius consistently leads to lower RMSE of
  the target position estimates (for any of the considered
  algorithms), and of the sensor position estimates (only for
  SLAT). This behavior is expected since a higher radius allows more
  sensors to perform measurements, which are then used to update both
  the target's and the sensors' positions. However, recall that the
  sensing radius can be only increased up to a limit defined by the
  maximum sensing threshold for which the measurement model is
  valid. We also note that the difference between the tracking and
  localization algorithms is decreasing with an increasing sensing
  radius. This can be explained by the fact that the IMU provides a
  smaller proportion of the total information when more sensors
  provide measurements.

\item \textit{Effect of varying the number of sensors
    (Figs. \ref{fig:rmse-param}b and \ref{fig:rmse-param}e):}
  Increasing the number of deployed sensors will obviously lead to
  better performance of all considered algorithms. Thanks to the
  measurements from the IMU, the sensitivity is not significant for
  the target position estimates of the tracking and SLAT
  algorithms. Regarding the sensors' SLAT estimates, the sensitivity
  is extremely low because each sensor receives the same number of
  messages from the target regardless of the number of sensors (see
  Fig. \ref{fig: slat-mrf}). Nevertheless, these messages are more
  informative since they are functions of improved target's
  beliefs. Finally, we can conclude that the SLAT algorithm can be
  used with a lower density of deployed sensors than the corresponding
  localization/tracking algorithms (e.g., if an RMSE of 3 m is
  acceptable, we need 10 sensors in case of SLAT, 15 sensors in case
  of tracking, and 22 sensors in case of localization). This is
  especially important if the sensors are expensive.

\item \textit{Effect of varying the standard deviation of the sensors'
    positions (Figs. \ref{fig:rmse-param}c
    Fig. \ref{fig:rmse-param}f):} In all analyzed cases, the RMSE is
  nearly a linear function of $\sigma_s$. However, the most important
  difference between the algorithms is the slope of the curve, which
  is the lowest in case of SLAT for both the sensor and the target
  position estimates. That basically means that SLAT is the most
  useful in scenarios in which the sensors are very imprecisely
  deployed, or if most of the sensors drop far from their original
  positions (which may happen in the aftermaths of accidents).  \ei

\begin{figure*}[!t]
\centerline{
\subfloat[]{\includegraphics[width=0.44\textwidth]{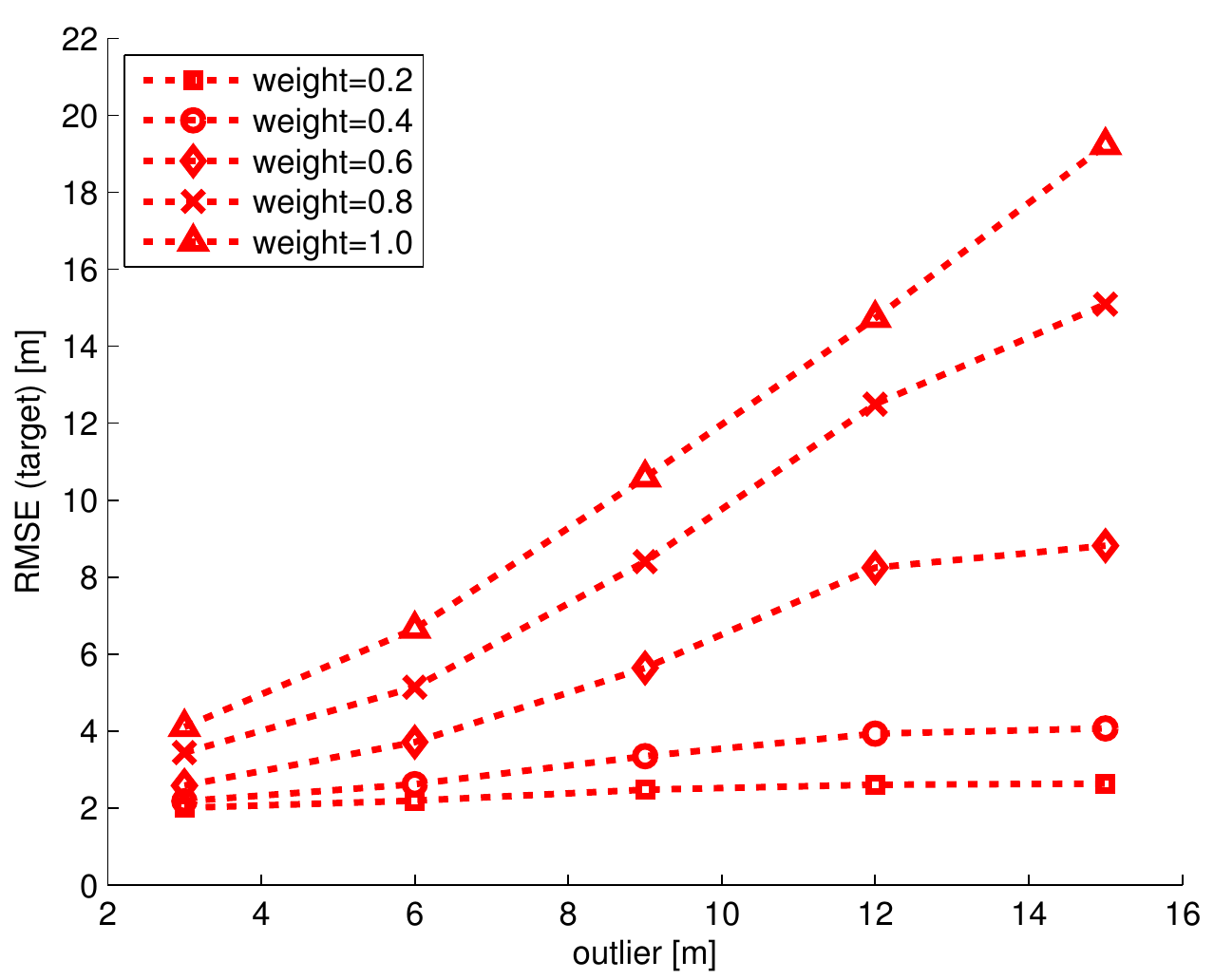}}\label{fig:rmse-outl-targ} 
\hfil
\subfloat[]{\includegraphics[width=0.44\textwidth]{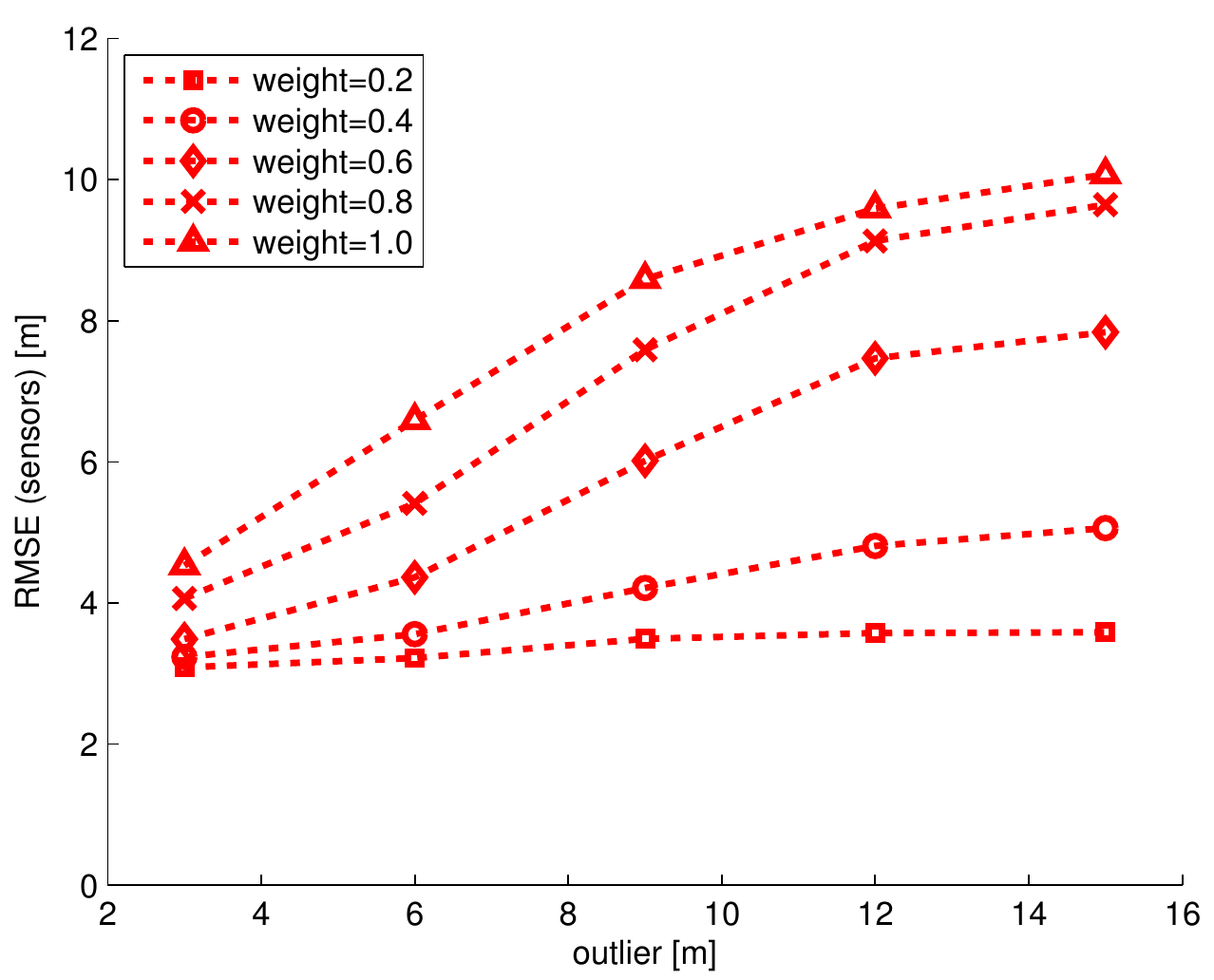}}\label{fig:rmse-outl-sens}
}
\caption{RMSE of the SLAT algorithm as a function of parameters of the
  contamination noise for (a) the target's position error, and (b) the
  sensors' position error.}
\label{fig:rmse-outl}
\end{figure*}

Finally, we analyze the robustness of the SLAT algorithm against
outliers in the distance measurements. Therefore, we assume that the
measured distance is contaminated with an outlier $d_\mathrm{o}$, with
weight $P_\mathrm{o}$. In other words, the contaminated measured
distance is given by $d_{\mathrm{o},t,n}=d_{t,n}+n_\mathrm{o}$ where
$n_\mathrm{o}$ is contamination noise distributed as:
\be\label{eq:contam} n_\mathrm{o} \sim
(1-P_\mathrm{o})\delta(n_\mathrm{o})+P_\mathrm{o}\delta(n_\mathrm{o}-d_\mathrm{o}).
\ee 
Note that the presence of these outliers is unknown to the SLAT
algorithm, so it is not part of the model in
\eqref{eq:dist-toa}. These outliers may be caused by additional NLOS conditions, 
interference, or other sources of errors.
According to the results, shown in
Fig. \ref{fig:rmse-outl}, we find the following: i) higher values of
$P_\mathrm{o}$ and/or $d_\mathrm{o}$ would lead to increased RMSE, ii)
small value of $P_\mathrm{o}$ would not lead to performance
degradation regardless of the value of $d_\mathrm{o}$, iii) the sensor
position estimates are less sensitive to outliers than the target
position estimates, and iv) for $P_\mathrm{o}>0.4$ and large values of
$d_\mathrm{o}$, the sensor position estimates are worse than the
corresponding tracking/localization estimates, in which the RMSE is
equal to 5.9 m (see Fig. \ref{fig:rmse-param}e) even in the presence
of outliers. In principle, the algorithm is very robust against
outliers, especially for small weights $P_\mathrm{o}$, and this is
achieved thanks to the uniformly distributed tail in the distribution
of the measurement noise (see \eqref{eq:measure-slat}). The unique
problem is that SLAT should not be used if these weights are large,
but this situation should not be expected in reality. If the
most of the links are contaminated with an outlier (this situation
can be detected by comparing tracking and SLAT), it is necessary to
perform re-calibration, as explained in Section \ref{subsec:issues}.

\subsection{Experimental Example}\label{subsec:exper-example}

We consider a small part of the CANMET mine \cite{Chehri2009} to test the performance of our algorithm. The tunnel is divided into $N_c=14$ (non-equal) cells, and in each of them there is one sensor (i.e., $N_s=N_c$). The sensors' priors are given by ${p_{n,0}}({{\bf{z}}_{n,0}}) = {\cal N}({{\bf{z}}_{n,0}};{{\bf{l}}_n},{\Sigma _S})$ where ${{\bf{l}}_n}$ ($n=1,\ldots,N_s$) is the expected location of the sensors, and ${\Sigma _S} = {\rm{diag}}(25\,{\rm{m^2}},\,25\,{\rm{m^2}},\,0\,)$. There are $N_T=30$ time slots, the sampling interval is $T_s=1$ s, and the quantization noise parameter is $D=6$ m. Taking the results from \cite{Chehri2009}, the measurement noise follows a Gaussian distribution for LOS, and a Weibull distribution for NLOS. More details about the considered scenario, and all other parameters, are available in \cite{Savic2013fusion}.

\begin{figure*}[!t]
\centerline{
\subfloat[]{\includegraphics[width=0.44\textwidth]{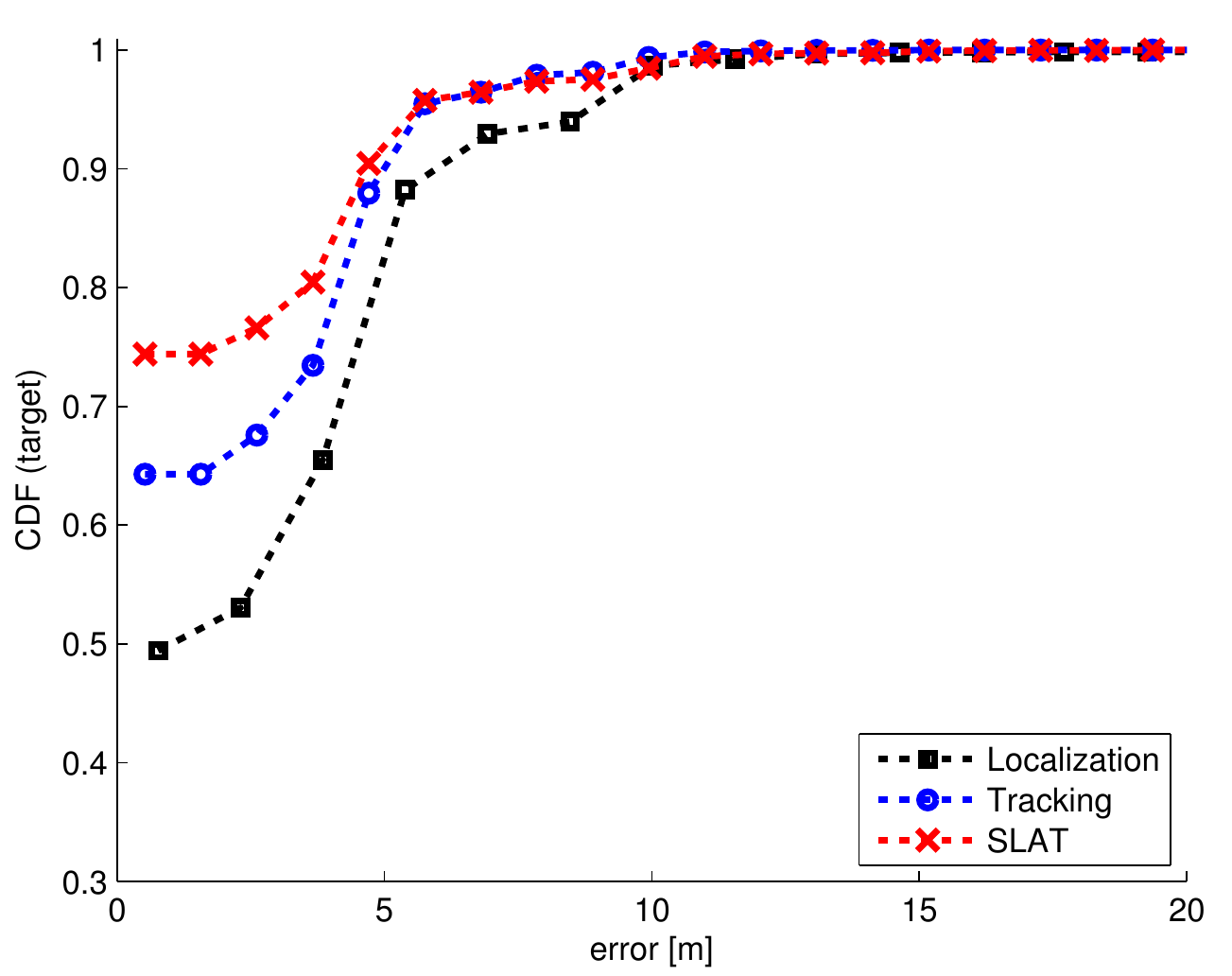}\label{fig:cdf-targ-canmet}}
\hfil
\subfloat[]{\includegraphics[width=0.44\textwidth]{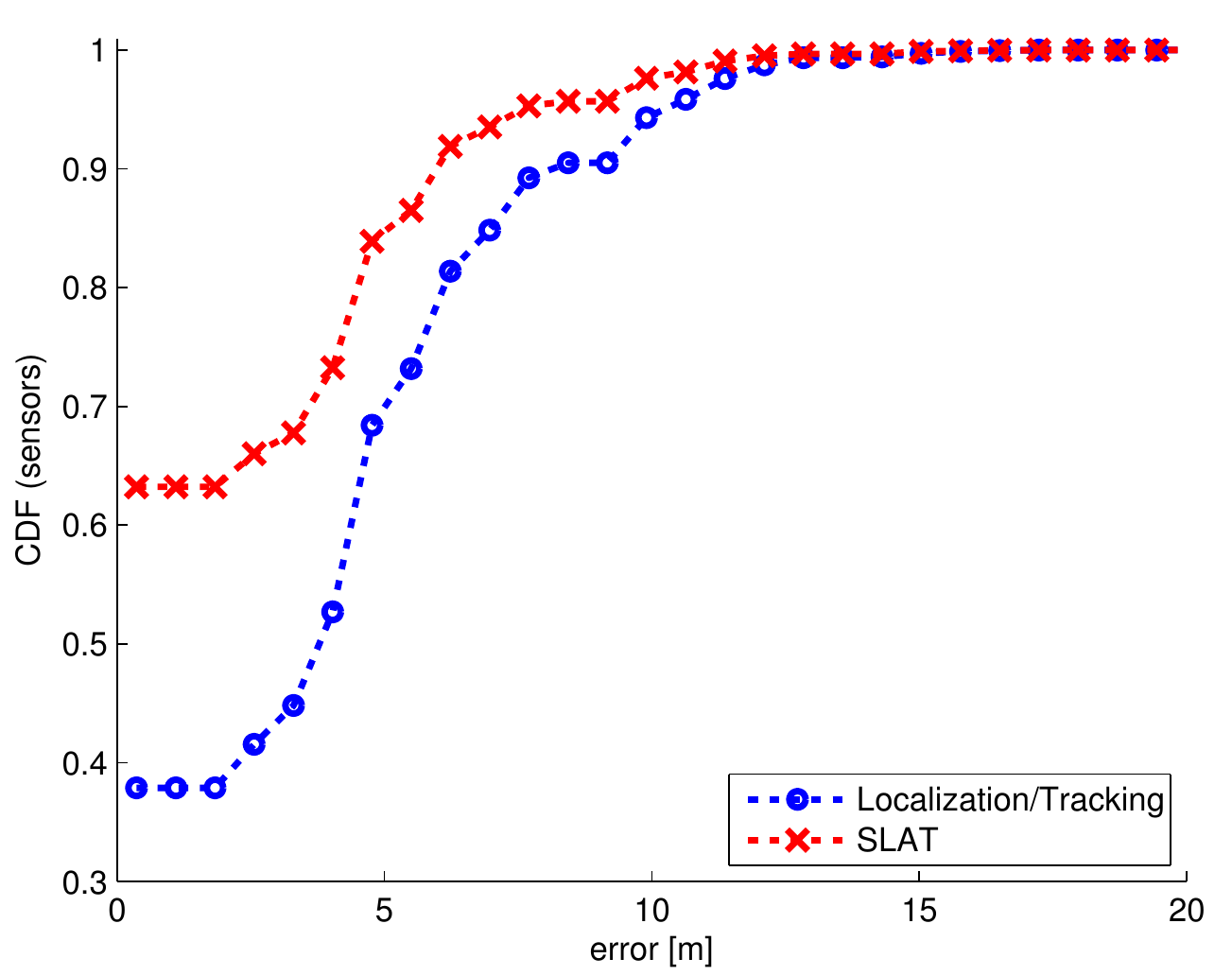}\label{fig:cdf-sens-canmet}}
}
\caption{CDF of (a) the target's position error, and (b) the sensors' position error. The results are based on models from the CANMET mine.}
\label{fig:cdf-error-canmet}
\end{figure*}

We analyze the CDF of the target's and the sensors' position error, shown in Fig. \ref{fig:cdf-error-canmet}. As we can see, SLAT is consistently (at any percentile) more accurate than corresponding tracking and localization algorithms, which is consistent with our results based on ray-tracing (Fig. \ref{fig:cdf-error}). One minor difference is that CDF is smoother than in Fig. 6, which is caused by variable cell sizes.

\subsection{Other solutions in literature}\label{subsec:disc}

Although not for the same type of the environment, there are other solutions in literature that try to address a similar problem. They are mainly based on fingerprinting techniques \cite{Nerguizian2006,Bshara2010,Ergen2014}, and SLAT algorithms over continuous space \cite{Taylor2006,Teng2012,Kantas2012}. In general, fingerprinting is able to outperform our proposed SLAT method, assuming that enough fingerprints are available. However, this approach would require an exhaustive calibration that sometimes may be infeasible. An additional problem of fingerprinting is that sensors may fall from the walls during the training phase (which have to be repeated very frequently), making a subset of the fingerprints invalid. Regarding other SLAT algorithms, to the best of our knowledge, none of them are adapted to non-Gaussian measurement models and for confined areas. For example, the SLAT methods in \cite{Taylor2006,Teng2012,Kantas2012} provide the posterior in Gaussian form, and use an unrestricted continuous space. Consequently, they would either fail in the presence of high levels of NLOS bias or provide invalid estimates (e.g., behind the walls). 
\\
\section{Conclusions and Future Work}\label{sec:conc}
We presented a novel approach to target tracking in confined
  environments in the presence of uncertain sensors positions. It is
  based on a SLAT-BP algorithm, which can simultaneously, and in
  real-time, refine the sensors' positions and track the target. This
  algorithm can: i) efficiently solve high-dimensional
  problems, ii) handle all non-Gaussian uncertainties, and iii)
  provide only valid position estimates thanks to the use of cells
  with predefined positions. According to our simulation results, both
  the sensor and the target position estimates are improved even after
  a very short tracking period. Moreover, SLAT-BP can be used with a
  very low density of deployed sensors, and can keep performing well
  even if most of the sensors drop far from their original
  position. Finally, the algorithm is also very robust against
  outliers as long as they appear with reasonable probabilities. 

By no means the study in this paper provides the solutions for
  all problems in confined environments. There remain many open
  directions for  future work. First, it would be useful to provide
  a method that learns and adapts the ranging distribution online. As a
  result, the SLAT algorithm can be re-used for a wide variety of
  environments. Second, a distributed implementation of the SLAT
  algorithm, in which the sensors estimate the target's position by
  cooperating only with the neighboring sensors, may be useful to
  increase the scalability and robustness. Finally, cooperative
  infrastructure-free self-localization algorithm would be of high
  interest for search-and-rescue operations.

\section*{Acknowledgment}

The authors would like to thank Sasit Chuasomboon for helping with the
ray-tracing simulations. We also thank Peter Stenumgaard (FOI) and
Fredrik Gustafsson (Link\"oping Univ.) for their valuable comments and
suggestions.

\footnotesize
\bibliographystyle{ieeetr} 
\bibliography{../../../../Publications/vs-refs,../../../../Publications/others}

\begin{IEEEbiography} [{\includegraphics[width=1in,height=1.25in,clip,keepaspectratio]{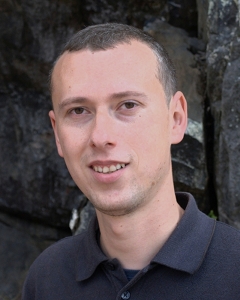}}] {Vladimir~Savic}
received the Dipl.Ing. degree in electrical engineering from the University of Belgrade, Serbia, in 2006, and the M.Sc.
and Ph.D. degrees in communications technologies and systems from the Universidad Politecnica de Madrid, Spain, in 2009 and 2012, respectively. He was a Digital IC Design Engineer with Elsys Eastern Europe, Belgrade, from 2006 to 2008. From 2008 to 2012, he was a Research Assistant with the Signal Processing Applications Group, Universidad Politecnica de Madrid. He spent three months as a Visiting Researcher at the Stony Brook University, NY, and four months at the Chalmers University of Technology, Gothenburg, Sweden. In 2012, he joined the Communication Systems (CommSys) Division, Link\"{o}ping University, Sweden, as a Postdoctoral Researcher. His research interests include localization and tracking, wireless channel modeling, Bayesian inference, and distributed and cooperative signal processing in wireless networks.
\end{IEEEbiography}

\begin{IEEEbiography} [{\includegraphics[width=1in,height=1.25in,clip,keepaspectratio]{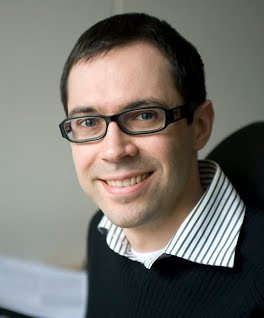}}] {Henk~Wymeersch}
(S'99, M'05) received the Ph.D. degree in Electrical Engineering/Applied Sciences in 2005 from Ghent University, Belgium. He is currently an Associate Professor with the Department of Signals and Systems at Chalmers University of Technology, Sweden. Prior to joining Chalmers, he was a Postdoctoral Associate with the Laboratory for Information and Decision Systems (LIDS) at the Massachusetts Institute of Technology (MIT). He served as Associate Editor for \emph{IEEE Communication Letters} (2009--2013), \emph{IEEE Trans. on Wireless Communications} (2013--present), and \emph{Trans. on Emerging Telecommunications Technologies} (2011--present). 
\end{IEEEbiography}

\begin{IEEEbiography} [{\includegraphics[width=1in,height=1.25in,clip,keepaspectratio]{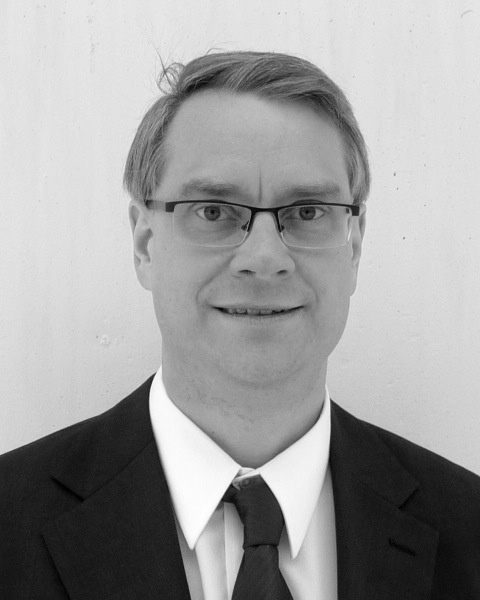}}] {Erik G. Larsson}
received his Ph.D. degree from Uppsala University, Sweden, in 2002.  Since 2007, he is Professor and Head of the Division
for Communication Systems in the Department of Electrical Engineering (ISY) at Link\"oping University (LiU) in Link\"oping, Sweden. He has previously been Associate Professor (Docent) at the Royal Institute of Technology (KTH) in Stockholm, Sweden, and Assistant Professor at the University of Florida and the George Washington University, USA.

His main professional interests are within the areas of wireless communications and signal processing. He has published some 100 journal papers on these topics, he is co-author of the textbook \emph{Space-Time Block Coding for Wireless Communications} (Cambridge Univ. Press, 2003) and he holds more than 10 issued and many pending patents on wireless technology. He has served as Associate Editor for several major journals, including the \emph{IEEE Trans. on Communications} (2010-2014) and \emph{IEEE Trans. on Signal Processing} (2006-2010). He serves as  chair of the IEEE Signal Processing Society SPCOM technical committee in 2015--2016. He also serves as chair of the steering committee for the \emph{IEEE Wireless Communications Letters} in 2014--2015.  He is active in conference organization, for example as General Chair of the Asilomar Conference on Signals, Systems and Computers in 2015 (he was Technical Chair in 2012).  He received the   \emph{IEEE Signal Processing Magazine} Best Column Award twice, in 2012 and 2014.
\end{IEEEbiography}

\end{document}